\documentclass[12pt,preprint]{aastex}

\shorttitle{CORRELATION BETWEEN GALAXY MERGERS AND LUMINOUS AGN}
\shortauthors{Hong\&Im et al.}

\begin{document}


\title{CORRELATION BETWEEN GALAXY MERGERS AND LUMINOUS ACTIVE GALACTIC NUCLEI}


\author{Jueun Hong\altaffilmark{1,2,6}, Myungshin Im\altaffilmark{1,6}, 
Minjin Kim\altaffilmark{3}, \& Luis C. Ho\altaffilmark{4,5}}
\affil{\altaffilmark{1}Center for the Exploration of the Origin of the Universe (CEOU), 
Astronomy Program, Department of Physics \& Astronomy, Seoul National University, 
Seoul, Republic of Korea}
\affil{\altaffilmark{2} Donga Science Co., Ltd., 
7th Floor, 109, Cheongpa-ro, Yongsan-gu, Seoul, 140-877, Korea}
\affil{\altaffilmark{3}Korea Astronomy and Space Science Institute, Daejeon 305-348, Korea}
\affil{\altaffilmark{4}{Kavli Institute for Astronomy and Astrophysics,
Peking University, Beijing 100871, China}}
\affil{\altaffilmark{5}The Observatories of the Carnegie Institution for Science, 813 Santa Barbara Street, Pasadena, CA 91101, USA}

\email{\altaffilmark{6}jueunhong@astro.snu.ac.kr, mim@astro.snu.ac.kr}




\begin{abstract}
 It is not yet clear what triggers the activity of active galactic nuclei (AGNs), but 
 galaxy merging has been suspected to be one of the main mechanisms fuelling the activity. 
 Using deep optical images taken at various ground-based telescopes, 
 we investigate the fraction of galaxy mergers in 39 luminous AGNs (M$_{R}\, \lesssim$ -22.6 mag) 
 at $z \leq$ 0.3 (a median redshift of 0.155), of which the host galaxies are generally considered as early-type  
 galaxies. 
 Through visual inspection of the images, we find that 17 of 39 AGN host galaxies (43.6\%) show 
 the evidence for current or past mergers like tidal tails, shells, and disturbed morphology.
 In order to see if this fraction is abnormally high, we also examined the merging fraction of normal
 early-type galaxies in the Sloan Digital Sky Survey (SDSS) Strip 82 data (a median redshift of 0.04),
 of which the surface-brightness limit is comparable to our imaging data. To correct for the effects related to the redshift difference 
 of the two samples, we performed an image simulation by putting a bright point source as an 
 artificial AGN in the images of SDSS early-type galaxies and placing them onto the redshifts of AGNs. 
  The merging fraction in this realistic sample of simulated AGNs is only $\sim 5 - 15\%$ 
  ($1/4$ to $1/8$ of that of real AGNs). 
  Our result strongly suggests that luminous AGN activity is associated with galaxy merging.
\end{abstract}


\section{INTRODUCTION}

 An active galactic nucleus (AGN) is an energetic core of a galaxy whose energy 
 comes from accretion of surrounding matter by a supermassive black hole (SMBH) 
 that resides at the central part of the galaxy.
In order for an SMBH to shine as an AGN, it needs a supply of gas to fuel its activity. Two main mechanisms have been suggested to trigger AGN activity: an internal mechanism through 
a dynamical instability inside a galaxy and an external mechanism through galaxy-galaxy interaction or merging. However, it is not yet clear which one is the dominant mechanism,
  even after many observational studies have been carried out.

  The internal mechanism is such that a gas inflow to the central part occurs as a result of 
 instability in the internal structure of a galaxy.
  For example, a galaxy bar can move gas from the outer regions of a galaxy into its center, and then the gas inflow 
  can trigger the AGN phase 

\citep{Lynden-Bell(1979), Sellwood(1981), van Albada1981, Combes1985, Pfenniger1991, Heller1994, Bournaud2002, Athanassoula(2003), sakamoto}.
   Studies of AGNs with moderate X-ray luminosity, especially the studies targeting distant AGNs, 
  find that most of the AGN hosts are disk galaxies and do not show evidence of a major merger, 
  supporting the idea that an internal mechanism triggers 
  the AGN activity and the SMBH growth \citep{Schawinski(2012), Schawinski(2011), cisternas, Kocevski, simmons}.
   In the local universe, the fraction of Seyfert galaxies is found to be high in barred galaxies 
  in the Sloan Digital Sky Survey (SDSS) data, 
  suggesting a close connection between bar and AGN activity \citep{oh}, but
  studies exist that point to an opposite conclusion \citep{lee}.

   In contrary, the external mechanism is represented by galaxy interaction or merging. 
   In such a mechanism, gas infall during a major merging triggers AGN activity
 \citep[e.g.,][]{barnes, mihos1996, hopkins2005, springel}. 
  There are a number of observational results that support this idea. Studies of galaxy pairs or galaxies in interaction find that the AGN fraction increases  
  in such systems. \citep{Ellison2011, Silverman, Bessiere}.
   Ultraluminous infrared galaxies (ULIRGs) are suggested to harbor AGNs, and many ULIRGs are merging systems \citep[e.g.,][]{sanders1988_1, sanders1988_2}. Binary SMBHs in some 
  AGNs demonstrate that two or more SMBHs can merge into one SMBH \citep{Koss, Komossa}.
  After all, many AGN host galaxies are found to be elliptical galaxies, which do not possess
  bars or disk instabilities. 

   One promising way to investigate the AGN and merger connection is to study objects 
  with merging features. When two galaxies with comparable mass merge, 
   the merging produces an early-type galaxy \citep{toomre, barnes1992}, leaving a trace 
  of the past merging activity in the form of tidal tails, shells, and dust lanes \citep{Quinn, Hernquist, feldmann}.
    In support of this theoretical expectation, very deep imaging of early-type galaxies
  find merging features in many cases \citep[15$\%-80\%$, depending on the depth of the image;][]{vanDokkum, Tal, kaviraj, kimtaehyun, Sheen}.
   The tidal features remain for about 1-2 Gyr after the merging \citep{feldmann},
  which overlaps with the period of AGN activity that occurs in some models 
  at the same timescale of roughly 1 Gyr after  
  the merging \citep{hopkins2005}.
   Therefore, if major merging triggers AGN activity, we expect 
  the following to be seen: (1) the AGN host will be an early-type galaxy or a bulge-dominated
  system, (2) merging features such as tidal tails and shells should be observable 
  preferentially in AGNs than in nonactive early-type galaxies, and (3) AGNs should be found
  more in spheroidal postmergers (1-2 Gyr after the merging) than in merging systems 
  at a very early stage.
  
   Interestingly, previous searches for merging features in AGN host galaxies have seen mixed 
   results. \cite{bennert} found that four out of five low-redshift QSO hosts show obvious merging features 
   from deep Hubble Space Telescope (HST) imaging data. A more extensive sample of 69 QSO hosts (0.142 $\leq z \leq$ 0.198)
   was imaged by a ground-based telescope, and it was found that 
  60\% of the sample shows signs of interaction \citep{letawe}. 
  \cite{Koss} studied 
  the fraction of mergers in the Swift BAT-selected AGNs and found that the fraction  
  of BAT AGNs showing evidence of merger is much higher than the fraction in a sample of inactive
  galaxies (18\% versus 1\%). For a special class of dust-reddened AGNs, \cite{Urrutia}
 also found a high fraction of AGN hosts with evidence for ongoing or recent
 interaction (85\%). \cite{Almeida} investigated the brightness of merging features in
  radio galaxies at $0.05 < z < 0.7$ and found that the merging features of radio-loud active
  galaxies are 2 mag brighter than those of quiescent galaxies. 
\cite{Carpineti} found that about 70\% of spheroidal postmergers \textemdash spheroidal galaxies with merging features \textemdash are Seyferts or LINERs.
By comparing this fraction to ongoing merging systems, they suggested that 
   the AGN phase becomes more dominant only in the very final stage of the merging. 
   On the other hand, an early study of \cite{Bahcall} found 
  that only 15\% of AGN hosts show a clear indication of interaction, although
 they also noted a large fraction AGN hosts with close, projected companions. However, 
 the more recent studies of Karouzos et al. (2014a, 2014b) show that there is 
 no enhancement of close neighbors around
 X-ray or radio-selected AGNs with respect to field galaxies with properties similar to 
 the AGN hosts. 
  The aforementioned studies supporting the internal AGN triggering mechanism all find 
 that the fractions of both early-type host galaxies and merging systems are low
 in distant AGNs.

   These pieces of seemingly contradictory results can be resolved if AGN activities
  are triggered by different mechanisms at different levels.
  There are theoretical studies that suggest such an idea, with the external mechanism
 preferentially triggering luminous AGN activity \citep{hopkins2009, Draper}. 
  From an observational point of view, the studies favoring the merging mechanism are mostly based
  on QSOs (bright AGNs) and early-type  host galaxies.
  In comparison, the observational studies favoring the internal mechanism are
  focused more on less-luminous AGNs or late-type host galaxies. 
\cite{Treister} suggest that
  the fraction of merging systems in AGNs is a strong function of AGN luminosity, 
  with merging activity being stronger in more luminous AGNs (AGNs with a large bolometric luminosity).
  Their sample, however, is rather heterogeneous, making it difficult to draw a firm conclusion on 
  this issue. Furthermore, a more recent study excludes a strong connection between AGNs and mergers over a 
  wide range of X-ray luminosity ($10^{41} < L\mathrm{(X-ray)erg^{-1}sec^{-1}} < 10^{44.5}$), 
  based on their investigation of AGN host galaxies at $0 < z < 3$ in the Chandra Deep 
  Field south, including a careful treatment of a control sample that is compared against the morphology of
  the AGN host galaxies \citep{Villforth}.
  A similar conclusion was reached by \cite{Georgakakis} too.

    Although researchers do not yet agree on the significance of merging for AGN triggering
 activity, it is known that host galaxies of luminous AGNs are predominantly early-type galaxies
 \citep{mclure, dunlop}, i.e., products of major merging, so it is natural to consider merging as 
  a key AGN triggering mechanism in luminous AGNs. 
  The definition of ``luminous" is a bit ambiguous, but the luminosity cut around
  $M_{R} \sim -23$mag qualifies as ``luminous" for the central point source. While previous studies have investigated the connection between merging features and early-type hosts, the studies are rather limited
  in sample size \citep[e.g.,][]{bennert} or they include a heterogeneous sample of host galaxy types
  and are not geared to detect the faintest features \citep[e.g.,][]{Georgakakis, cisternas, Schawinski(2011), Kocevski, Treister}. 
  While deriving the fraction of merging
  signatures in AGNs can be useful, interpreting the result as a sign of merger-triggered AGN activity
  needs an additional step of comparison with an ordinary galaxy sample because many normal early-type 
  galaxies are found to have merging features. However, such a comparison is a complicated
  process because the surface-brightness limits often differ between an AGN sample and a normal galaxy sample,
  and the bright point sources in AGNs can make it more challenging to detect merging features.

   In order to address the specific question of whether luminous AGNs are triggered by merging activity,   
 we study the fraction of luminous AGNs showing merging features and compare it to the 
 fraction of normal galaxies with merging features.
  Because the light from luminous AGNs can contaminate the outer regions of their host galaxies, 
 deep images of luminous AGN are obtained to reveal merging features.
  For a similar reason, we use a control sample of normal early-type galaxies selected from 
  the SDSS Stripe 82 data, which reaches a surface-brightness limit of $\mu_{r} \sim 27$ mag arcsec$^{-2}$ over a 1$\arcsec$ $\times$ 1$\arcsec$ area at rms noise of 1$\sigma$ of the background regions of the image\footnote{We will use this definition of SB limit throughout the paper. Even though less rigorous than the definition of the SB limits in \cite{Li}, \cite{Ho}, and \cite{Bakos}, this definition captures the depth of an image well in terms of how effectively we can identify faint, merging features.}.
  In addition, merging features are more difficult to find in AGN hosts than in the galaxies in the control sample because of the redshift differences and the contamination from the central point sources. Therefore, we simulated images of luminous AGNs using galaxy images in 
 the control sample so that the simulated AGNs contain a bright central point source and are placed 
 at redshifts similar to those of our AGN sample. This enables us to do a realistic comparison 
 between the control sample and the AGN sample. Finally, we compare the simulated AGNs versus the
 real AGNs, showing that merging features are much more easily found in  
 host galaxies of luminous AGNs than in the simulated AGNs.  
 Throughout the paper, we adopt a concordance cosmological parameters of 
 $H_0=70$ $\mathrm{km s^{-1}}$ $\mathrm{Mpc}$,  $\Omega_m=0.3$, and   $\Omega_\Lambda=0.7$
 (e.g., Im et al. 1997). 
 The physical quantities taken from the literature were converted to match 
 these cosmological parameters.

\section{AGN SAMPLE AND OBSERVATION}

\subsection{Sample Selection of AGNs} \label{sample selection _agn}

  We constructed two samples of luminous type-1 AGNs at $z < 0.3$, one that 
  we call as the ``base sample" and another that we call as the ``best sample," which is a subsample of the base sample. 
  The redshift cut of $z < 0.3$ is imposed because the features that we use to trace 
  the merging activity are overmatched by the bright point source
 of the AGN or become too faint to detect if a galaxy is too far away. 
  The definitions of ``luminous" are different for the two samples, and they are
  explained in the following. 
  
  The base sample comprises 39 luminous type-1 AGNs at $z < 0.3$.
  Here, the term ``luminous" is defined to be $M_{\mathrm{R}} \leq -22.6$ mag 
 ($M_{\mathrm{B}} < -22.0$ mag, or equivalently $M_{\mathrm{R}} < -22.6$ mag assuming B-R = 0.6), and ``type-1" AGNs are selected because they have an unambiguous spectroscopic signature (broad emission lines) for the existence of an AGN. 
  The objects are chosen with this luminosity cut 
 without separating the nuclear component from the host galaxy light, although we gave
 a preference to AGNs with bright nuclear components, such as Palomar Green (PG) quasars,
 during the sample selection process in order to select AGNs that are luminous 
 in the nuclear component.    
   Our analysis will focus on the base sample because the result from the best sample does not
 differ much from the result from the base sample, and the base sample provides better number statistics than does the best sample. The results from the best sample will be provided for the main results (Table 5 and Figures 18 and 19) 
 to demonstrate that the results from  both samples are identical.
 
   We also constructed a subsample of AGNs (33) based on the nuclear magnitude cut, 
  which we call as the ``best sample." 
   We have good estimates of nuclear magnitudes from either HST or Canada France Hawaii Telescope (CFHT) images for 
  a subsample of the base sample (59\% or 23 out of 39). For the remaining
  41\% of AGNs, we estimated their nuclear magnitudes 
  using a scaling relation between the black hole mass and the host galaxy bulge
  luminosity, as described in Section~\ref{mag_agn}.   
  For AGNs with good estimates of nuclear magnitudes (59\% of the base sample), 
  we find that most of them are dominated by a nuclear component 
  with  $m_{\mathrm{host}} - m_{\mathrm{Nuc}} \sim 1$ mag or 
  $m_{\mathrm{Nuc}} \sim m_{\mathrm{total}} + 0.36$ mag in median 
  (see Section \ref{simulation_control}). 
  The absolute magnitude cut of $M_{\mathrm{R}} < -22.6$ corresponds 
  roughly to $M_{\mathrm{Nuc}}(R) < -22.24$, assuming the above relation of 
   $m_{\mathrm{Nuc}} \sim m_{\mathrm{total}} + 0.36$ mag,
  and we adopt a slightly tighter nuclear magnitude cut of 
  $M_{\mathrm{Nuc}}(R) < -22.44$ to allow for a margin in the uncertainty in 
  the nuclear magnitude estimates. 
   The best sample will be used to support
  the validity of the results from the base sample.
  
\begin{figure}[p]
\includegraphics[width=14cm]{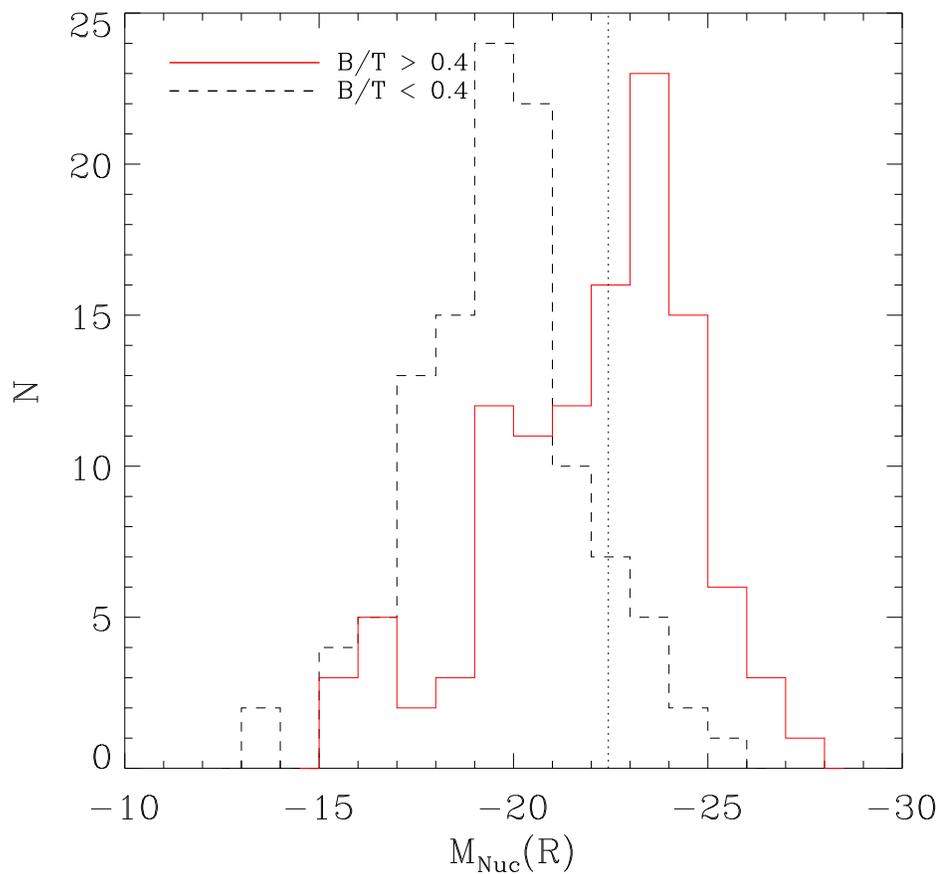}
\caption{Distribution of AGN nuclear magnitude in the $R$band from K15.
The red solid line shows the nuclear magnitude distribution of AGNs for which the host galaxy has $B/T \geq 0.4$.
The black dashed line shows the nuclear magnitude distribution of AGNs for which the host galaxy has $B/T < 0.4$. 
The black dotted line represents a cut at $M_{\mathrm{Nuc}} = -22.44$ mag. 
This figure shows that the host galaxies of AGNs with $M_{\mathrm{Nuc}}(R) < -22.44$ mag are mostly bulge-dominated galaxies.}.\label{kim}
\end{figure}

   Note that AGN host galaxies are mostly early-type galaxies 
 if their nuclear absolute magnitudes
 in the $R$ band ($M_{\mathrm{Nuc}}(R)$) are $M_{\mathrm{Nuc}}(R) \leq -22.44$ mag.
  Figure \ref{kim} shows the histogram of nuclear magnitudes of AGNs
 divided into two types---galaxies with $B/T > 0.4$ which we consider
 as early-type galaxies \citep{im2002, duhokim}, and galaxies with $B/T < 0.4$, which can be
 considered as late-type galaxies. Here, the data are taken from M. Kim et al. 
 (2015, in preparation, hereafter, K15)  
 where they decomposed the surface-brightness profile of AGNs with the HST archival
 images into host galaxies and the central point source components. 
 Figure \ref{kim} indicates that the host galaxies are dominated by early-type galaxies at 
 $M_{\mathrm{Nuc}}(R) < -22.44$ mag at the $\sim86.2\%$ level. 
 However, we did not impose the $B/T > 0.4$ cut in our AGN selection
  because the bulge absolute magnitude can be quite uncertain under the presence of a bright nuclear emission, with a typical error of $\pm0.2$ -- $0.4$ mag (Sanchez et al. 2004; Simmons et al. 2008; Kim et al. 2008b). The fraction of AGN hosts with $B/T < 0.4$ in our sample is small, anyway.
 
    The selection of AGNs was done in roughly two ways,  
    depending on the facilities used,
  because the observations were performed at one of three observatories 
 (Maidanak, McDonald, and Las Campanas) or the data were taken from the archive 
 (for two AGNs in our study).

   For observations at the Maidanak and McDonald observatories or for selecting objects from the archive,
  AGNs were selected from the Quasars and Active Galactic Nuclei catalog of \cite{veron}.
   Note that the catalog of \cite{veron} provides the absolute B-band magnitude only. In order to convert
  the B-band absolute magnitude to the R-band magnitude, 
  we assumed a quasar composite spectrum from \cite{vanden berk} which gives $B-R \sim 0.63$mag.
  This value is between 0.46 mag, calculated from the slope $\alpha_{\nu} \sim -1.03$ in $6005 \, \sim 7180\AA$ derived 
 by \cite{glikman}, and 0.71 mag, calculated from the slope $\alpha_{\nu} \sim -1.58$ in $4000 \, \sim 8555 \, \AA$ 
 derived by \cite{vanden berk}.
   We find 1,103 type-1 AGNs that satisfy the above sample-selection criteria. 
   The absolute magnitudes listed in the Veron-Cetty \& Veron catalog, however, may include  
 some amount of the light coming from the host galaxies.
  Therefore, we opted to observe the PG quasars among 
the possible targets that are observable during the assigned observing runs at
 the Maidanak and McDonald observatories because it is known that most of the PG quasars   
 have a strong nuclear component (e.g., Kim et al. 2008a). At the end, 
 the objects were selected mostly from the PG quasar catalog, 
 except for two AGNs for which the data were taken from the CFHT archive.
   The selected PG quasars were inspected 
 on the Digitized Sky Survey images, and none of them were found to be clearly dominated
 by a host galaxy component.
 Then, objects were randomly picked from a prepared target list  
for the observation.  
  Priority was given during the observations to AGNs with lower redshifts
 because morphological features become more difficult to identify for higher redshift objects 
 due to the surface brightness dimming and the resolution effects.
   In total, 18 AGNs of the sample and 14 AGNs of the best sample
  come from this selection.

  The sample observed with the DuPont 2.5-m telescope was taken from the \cite{kim2008a} 
  and K15 samples, which are made of luminous type-1 AGNs with HST images.
  Kim et al. (2008a) and K15 assembled an imaging data set from the {\it HST} archive of type-1 AGNs for which signal-to-noise ratio values of the chosen {\it HST} images are
  high enough to disentangle the host galaxy component from the bright nucleus 
  (Kim et al. 2008b). 
   Using the HST images and GALFIT, \cite{kim2008a} and K15 performed a 
  two-dimensional, multiple-component surface-brightness profile fitting of AGNs, 
 separating the nuclear component from the host galaxy and  
 decomposing the host galaxy surface-brightness profile into a bulge, a disk, and a bar if necessary.
  The  surface-brightness profile fitting provides structural parameters for the host galaxies.  
  Typical uncertainties of $B/T$ range from 0.05 to 0.2 (see also Ho \& Kim 2014).  
  Excluding three AGNs with host galaxies having a  
  clear signature of spiral arms, the number of AGNs that are observed by 
  the DuPont 2.5-m are 21 for the base sample
  and 19 for the best sample.

\begin{figure}[p]
\includegraphics[width=14cm]{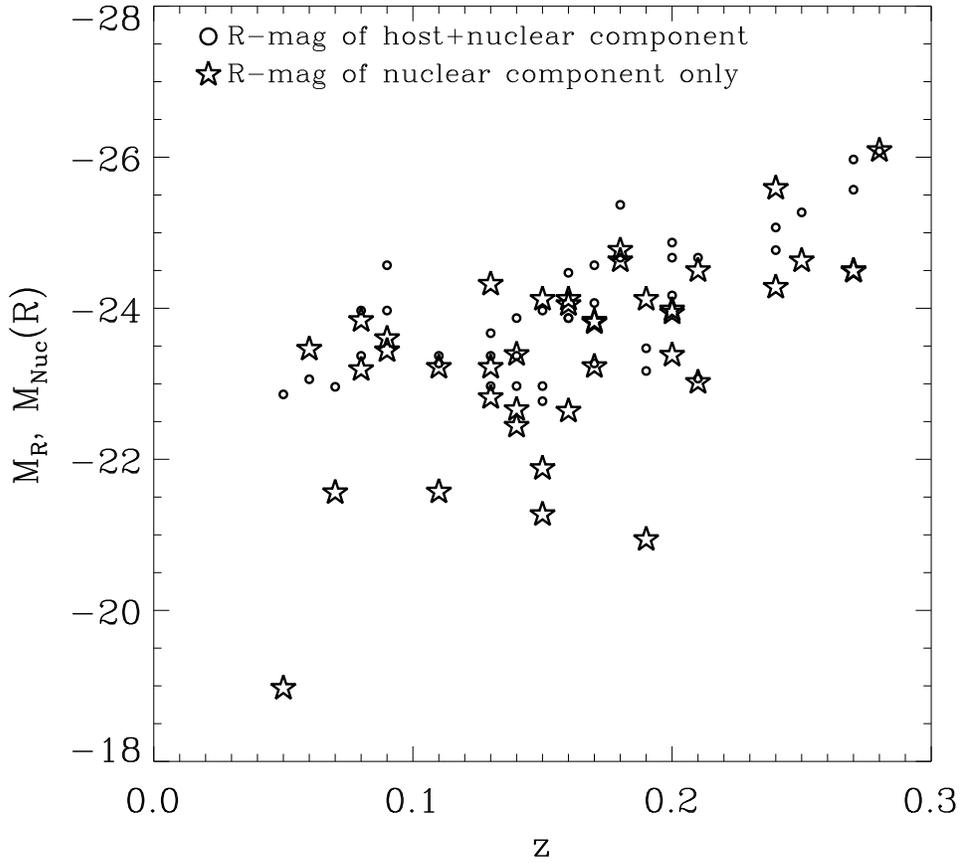}
\caption{Redshift versus absolute magnitude in the $R$ band of the host + nuclear component (small circles)
 and the nuclear component only ($M_{\mathrm{Nuc}}(R)$; stars) for 39 AGNs. See section \ref{mag_agn} for how we derived $M_{\mathrm{Nuc}}$.}\label{mr}
\end{figure}

  Table \ref{tbl-1} shows the list of AGNs in the base sample.
  Figure \ref{mr} shows the redshift versus the absolute magnitude of AGNs in the base sample in 
 both $M_{R}$ and $M_{\mathrm{Nuc}}(R)$. These AGNs 
 span a redshift range of $0.04 < z < 0.3$, with a median redshift of $z = 0.155$. 

    In summary, there are 39 and 33 type-1 AGNs at $z < 0.3$ in the base and the best 
  samples, respectively, where the best sample is based on a more rigorous sample-selection 
criteria than is the base sample, but at the expense of number statistics.
 Our base sample includes all 13 PG quasars with $M_R < -22.6$ mag, $z < 0.15$, 
and decl. $> 0$ degree and a total of 23 PG quasars.

\subsection{Observation and Data Reduction} \label{observation_1}
 In order to reveal faint, extended merging features around AGNs, 
 we obtained deep optical images using several facilities.
 We observed 10 AGNs using SNUCAM (Im et al.2010; FOV$=18\arcmin.1 \times 18 \arcmin.1$ and pixel scale$=0\farcs266$) on the 1.5m telescope at the Maidanak
 observatory in Uzbekistan from 2010 June to 2011 August, six AGNs using 
the Camera for QUasars in EArly uNiverse (CQUEAN; Park et al. 2012; Kim et al. 2011; Lim et al. 2013; FOV$=4\arcmin.8 \times 4\arcmin.8$ and pixel scale$=0\farcs281$)
 on the 2.1m telescope at the McDonald observatory in Texas from 2011 February to April, 
 and 21 AGNs using the SITe2K CCD camera of the 2.5m DuPont telescope 
 (FOV$=8\arcmin.85 \times 8\arcmin.85$ and pixel scale$=0\farcs259$)
 at the Las Campanas observatory from 2008 September to 2009 March. 
 Table \ref{tbl-1} gives the summary of the filter, the exposure time, the surface-brightness limit, and the observatory
 used for the observation of each target. The seeing ranges from $0\farcs8$ to 2$\arcsec$ with a typical value at around 1$\arcsec$.
 The adopted filters range from $V$ through $i$, depending on the observatory:
 The $V$ band for the SNUCAM data, the $i$ band for the CQUEAN data, 
 the $R$ band for the DuPont data,
 and the $i$ band or the $r$ band for CFHT archive data.
  We avoided using bluer filters (e.g., $B$) because tidal debris are suggested to be rather red
 \citep[e.g.,][]{feldmann}. Also, the choice of the filter was driven by the observational efficiency 
 to obtain deep images in minimal exposures. For example, CQUEAN at the McDonald observatory is
 optimized for red wavelengths and the $r$ band suffers from aberration near the CCD edge, 
 therefore the $i$ band was chosen. For the Maidanak observation, the balance between faint sky
 and the desire to avoid a fringing pattern forced us to choose the $V$ band. Although the chosen
 filter sets are nonuniform, this is not likely to cause serious systematic errors 
in our analysis.

 Since it is easier to identify merging features at the fainter surface-brightness limit ($\mu$), 
 we tried to reach the surface-brightness limit of $\sim$ 27 mag arcsec$^{-2}$, comparable to the depth of the control sample. 
 However, the resulting surface-brightness limit is a bit shallower than that of
 the control sample if the surface brightness dimming 
 is taken into account for objects at $z \gtrsim 0.1$.
To reach a limit of $\sim$ 27 mag arcsec$^{-2}$, we observed each target with a series of short exposure frames with exposure times of 60--180s, avoiding saturation of the central point source in the case of the Maidanak images, and a combination of short (a few hundred seconds) and long (1,000--2,000s) exposures for the DuPont images. These frames are combined to create deep images with total exposure times between 1 and 3 hr, 
 after properly rescaling images with different zero points and exposure times so that all of the frames, before stacking,
 share the same zero point.
 In some cases, clear merging features became visible even at a short exposure. 
 In such cases, the observation was stopped before reaching the desired surface-brightness limit 
(PG 0157+001 and PG 1613+658 at $\mu$ $\sim$ 26.11 and 26.18, respectively).
 The image of CTS J17.17 from the CFHT archive also shows a tidal tail at $\mu$ $\sim$ 26.0 mag arcsec$^{-2}$.

\begin{figure}[p]
\includegraphics[width=14cm]{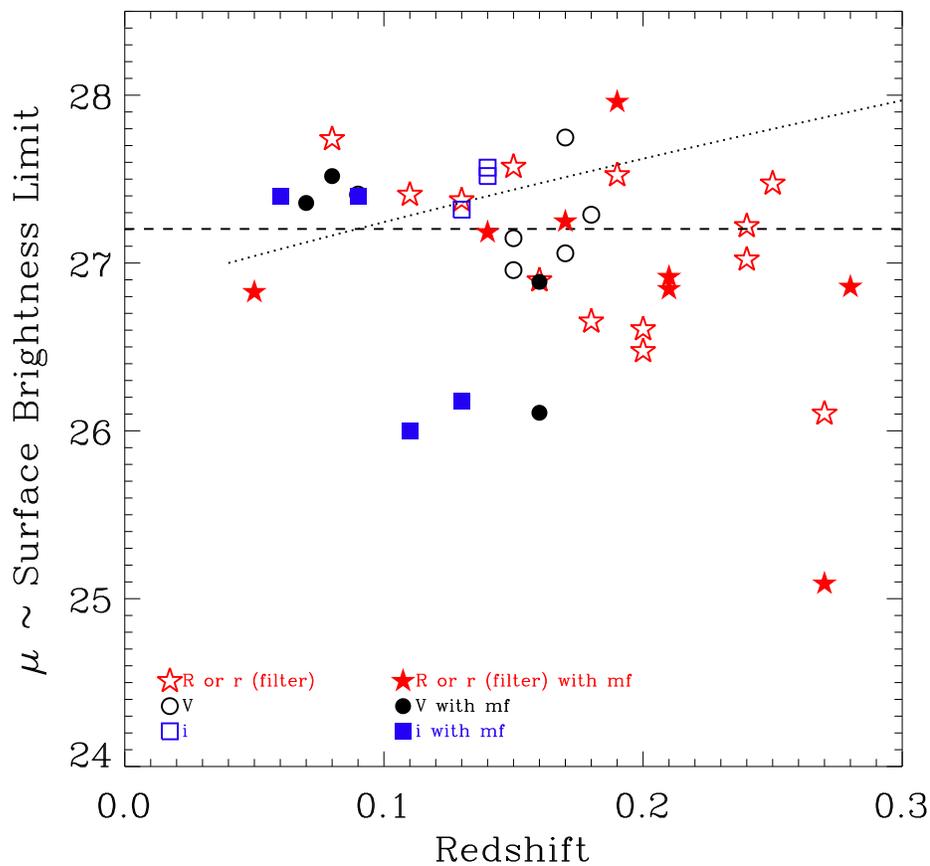}
\caption{Redshift versus observed surface-brightness limits of 39 AGNs. 
 The black dashed line is the median limit, 27.2 mag arcsec$^{-2}$ at 1$\sigma$.
 The red stars, the black circles, and the blue squares are the surface-brightness limits 
 in the $R$ or $r$, $V$, and $i$ filters, respectively. 
 The filled symbols denote AGNs that have merging features (see Section \ref{result}). 
 The dotted line indicates 
 the effective surface brightness limit
 of the control sample images 
 (See Section \ref{simulation_control}).
}\label{sb}
\end{figure}

   Figure \ref{sb} shows the redshift versus surface-brightness limit of the base sample. 
 Three different symbols represent the three different filters ($V$, $R$ or $r$, and $i$) used for the observation. 
 Note that the surface-brightness limit presented here is the ``observed'' surface-brightness limit. 
 The surface brightness dimming causes the rest-frame surface-brightness limit 
 of each object to be brighter by $\sim (1+z)^{4}$. 
 This introduces a bias in detecting merging features for an AGN at a higher
redshift: our AGN images are shallower than the control sample images for objects at a higher redshift. 
  To indicate this effect, we also plotted with the dotted line
 the ``effective" surface-brightness limit of the control sample. This is a
 surface-brightness limit of an image when galaxies in the image are
 assumed to be at a certain redshift. 
 For this line, we adopted 
 the surface-brightness limit of the SDSS Stripe 82 images 
 as 27 mag arcsec$^{-2}$ at their median redshift of 0.04. When 
 these images are assumed to be at a given redshift, the surface-brightness 
 limit at that redshift should be $27 + 10\, {\rm log}[(1+z)/1.04]$ mag arcsec$^{-2}$.
  As expected, we find that most of the AGN images are shallower than 
 the dotted line or comparable to the surface-brightness depth within 
 a few tenths of magnitude. 
 We will make a correction for this bias through simulations of AGNs (Section 3.2).

 Basic reduction of the images, such as the bias and the dark subtraction, and the flat-fielding was done
 using the standard IRAF tasks. Flat-field images were taken during twilight, 
 and these were used for flat fielding the images. 
 The dark correction was necessary only for the CQUEAN data where the dark current level is
 about 0.23 electrons/s/pixel \citep{Park} and not negligible in the data with a long exposure time.
 The reduced frames were registered to the position of the first frame 
 and stacked using the imcombine task of IRAF. Cosmic rays were rejected during the 
 stacking process, and the bad pixels and columns were fixed by the fixpix task of IRAF before 
 the registration of the frames. The photometry calibration was done by using standard stars observed
 for the SNUCAM and the DuPont data or by using the photometry information of 
 stars from SDSS in the vicinity of the targets for the CQUEAN data.  

\subsection{Black Hole Masses and Bulge, Nuclear, and Host Luminosities} \label{mag_agn}

 Because the merging fraction may depend on host-galaxy properties such as black mass or bulge 
 luminosity, we provide these quantities for our AGN sample in Table \ref{tbl-1}. 
 
 The black hole masses, $M_{\mathrm{BH}}$ in Table \ref{tbl-1} are taken from \cite{kim2008a} and 

\cite{vp}, where $M_{\mathrm{BH}}$ 
 values are estimated  
 by the reverberation mapping method for nine AGNs (marked as R in Table \ref{tbl-1}) and by single-epoch methods 
 for 28 AGNs. Here, the single-epoch method refers to the methods that estimate  $M_{\mathrm{BH}}$ values using spectra taken at a single epoch. In such methods,  the $M_{\mathrm{BH}}$ values are derived from a combination of the 5100 $\AA$ continuum luminosity and the line widths of broad emission lines (e.g., H$\beta$ or H$\alpha$; Vestergaard \& Peterson 2006).
The $M_{\mathrm{BH}}$ for the remaining two AGNs are derived from the bulge magnitude of the host, which will be explained
 in a later part of this section.
 We caution readers that the absolute values of $M_{\mathrm{BH}}$ can be off by a certain
amount. The $M_{\mathrm{BH}}$ estimates based on the virial method can be uncertain because they
depend on the scaling factor ($f$), which is sensitive to the geometry of the broad 
line region. Recently, Kormendy \& Ho (2013) showed that the correlation 
between the BH mass and the host galaxy mass in inactive galaxies is tight only for 
classical bulges and ellipticals. By comparing reverberation-mapped AGN to 
normal galaxies in the $M_{\rm BH}-\sigma_*$ plane only for classical bulges and 
ellipticals, Ho \& Kim (2014) recalibrated the scaling factor, which is approximately 
two times larger than the value used in this paper. Given the fact that our sample 
is hosted by early-type galaxies, it is very likely that bulges are either classical or 
ellipticals. Thus, it is possible that the BH mass is underestimated by a factor of two.

 The absolute magnitude of the bulge, $M_{\mathrm{Bul}}$, and the central point source, $M_{\mathrm{Nuc}}$, 
 are estimated in two ways.
 In the first method, $M_{\mathrm{Bul}}$ values are directly estimated from a two-dimensional surface-brightness fitting that
 takes into account the nuclear, bulge, and disk components. For AGNs with HST images, this is done
 in \cite{kim2008a} and K15, and we take the results from their work.
 Note that the absolute magnitudes of the nuclear component and the bulge component are derived with K corrections that assume a power-law spectral energy distribution for the nuclear component ($F_{\nu} \sim \nu^{\alpha}$ with $\alpha = -0.44$ blueward of $5,000 \, \AA$, and $-2.45$ otherwise; Vanden Berk et al. 2001) and an empirical template of elliptical galaxies for the bulge component. These absolute magnitudes include the correction for the Galactic extinction \citep{Schlegel}. For the AGNs with the ground-based images only, we attempted this procedure for two AGNs with the archival CFHT images 
because host galaxies are well resolved in these images. 
  We fitted the surface-brightness profile of AGNs using GALFIT \citep{Peng2002, Peng2010},
 assuming a model with three components, a point-spread function (PSF) for the central source and bulge and disk components
 for the host galaxy. The PSF was derived from a bright star in the vicinity of the AGN ($< 6\arcmin$).
  The disk component is represented by an exponential profile, and the bulge component is fitted with 
 a S\'{e}rsic profile \citep{sersic} as given below:

\begin{equation}
I(r)= I_e  exp[-b_n (r/r_e )^{1/n}-1], 
\end{equation}\label{eq:1}

 where $r_e$ is the effective radius, $I_e$ is the intensity at $r_e$, and $n$ is the Sersic index. During the fitting,
 the centers of each component were varied, and the background was fixed. 
 We also fitted the object in the vicinity of the target to improve the fitting result.
  When the scale length of one component is found to be too small ($< 1$ pixel), 
 we reduced the number of the models
 (PSF+bulge or PSF+disk) and used the deVaucouleurs profile for the bulge component.

\begin{figure}[p]
\includegraphics[width=16cm]{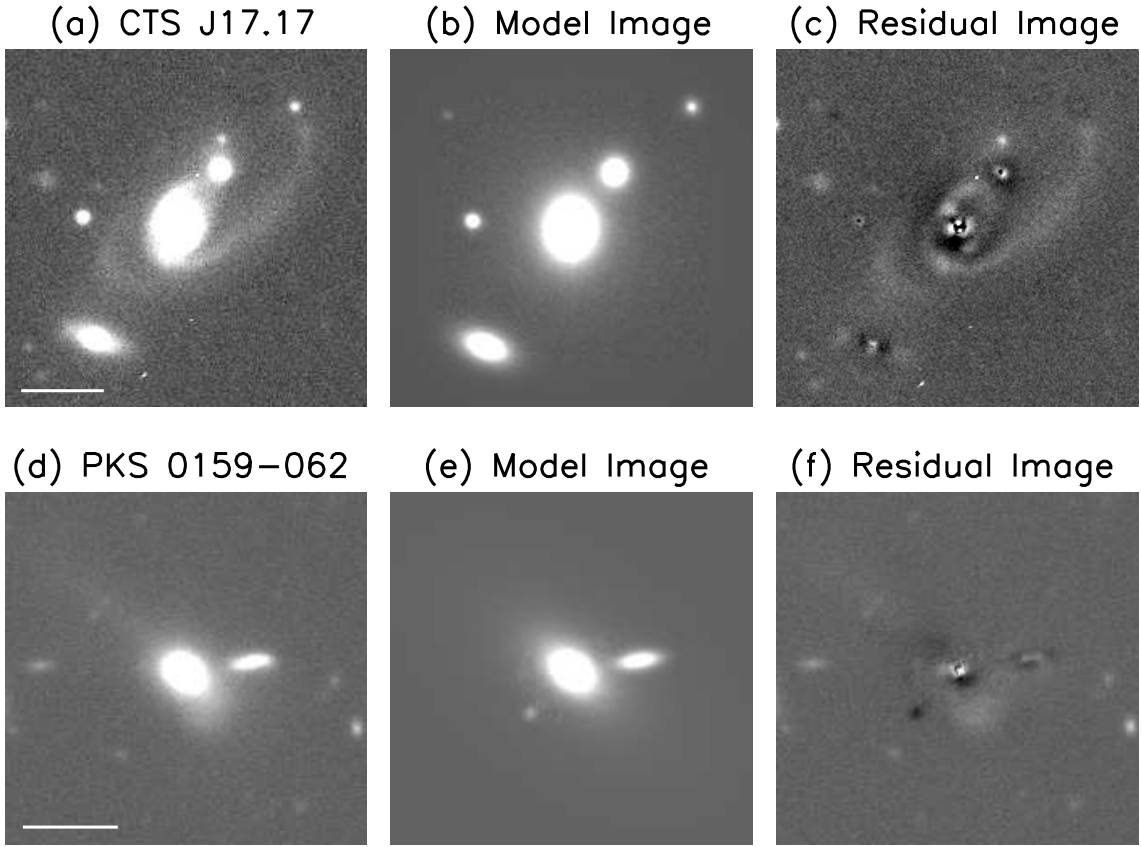}
\caption{CFHT images of two AGNs fitted by GALFIT: (a) original, (b) model, and (c) residual images of CTS J17.17,
 and (d) original, (e) model, and (f) residual images of PKS 0159-062.
The length of the horizontal bar in each panel is 10$\arcsec$. 
 The background of each image
 is subtracted, and the pixel values are square-rooted to bring out the low surface-brightness features. 
 Identical pixel value scales are adopted to plot images for the same object.
 }\label{galfit}
\end{figure}

 Figure \ref{galfit} shows the CFHT images of CTS J17.17 and PKS 0159-062 fitted by GALFIT. 
 The tidal tail and the jet-like features can be seen on not only 
 the residual images but also in the original images of the two AGNs. 
 Table \ref{tbl-2} shows the resultant fitting parameters. The absolute magnitudes are derived in the same manner as \cite{kim2008a}, which uses an empirical template of an E/S0 galaxy from \cite{calzetti} when deriving the K corrections.

 In the second method, $M_{\mathrm{Bul}}$ is estimated from $M_{\mathrm{BH}}$ using the correlation
 between $M_{\mathrm{Bul}}$ and $M_{\mathrm{BH}}$ of active galaxies \citep{kim2008a} as 

\begin{equation}
{\rm log} (M_{\mathrm{BH}}/M_{\odot}) = \alpha  + \beta M_{\mathrm{Bul}}(R),
\end{equation}\label{eq:ml}

 where $\alpha=-2.74$ and $\beta = -0.5$. Then $M_{\mathrm{Nuc}}$ is obtained by subtracting this $M_{\mathrm{Bul}}$ from the total luminosity of the AGN. This procedure assumes that $M_{\mathrm{Bul}} \sim M_{\mathrm{Host}}$ and thus may overestimate the nuclear luminosity if there is a significant disk component in the host and underestimate the nuclear luminosity if the object is an outlier of Eq. (2). We used this method for six AGNs for which neither 
 HST nor CFHT images were available. These are AGNs with $M_{\mathrm{BH}}$ values from \cite{vp}.
  We used the relation of Kim et al. (2008) instead of the $M_{\rm BH}$--$M_{\mathrm{Bul}}$ correlation of inactive galaxies because host galaxies of
 AGNs often show signs of a young stellar population (e.g., Kauffmann et al. 2003; Ho \& Kim 2014), leading to a somewhat smaller 
 stellar mass-to-light ratio in comparison to inactive galaxies. 

\begin{figure}[p]
\includegraphics[width=14cm]{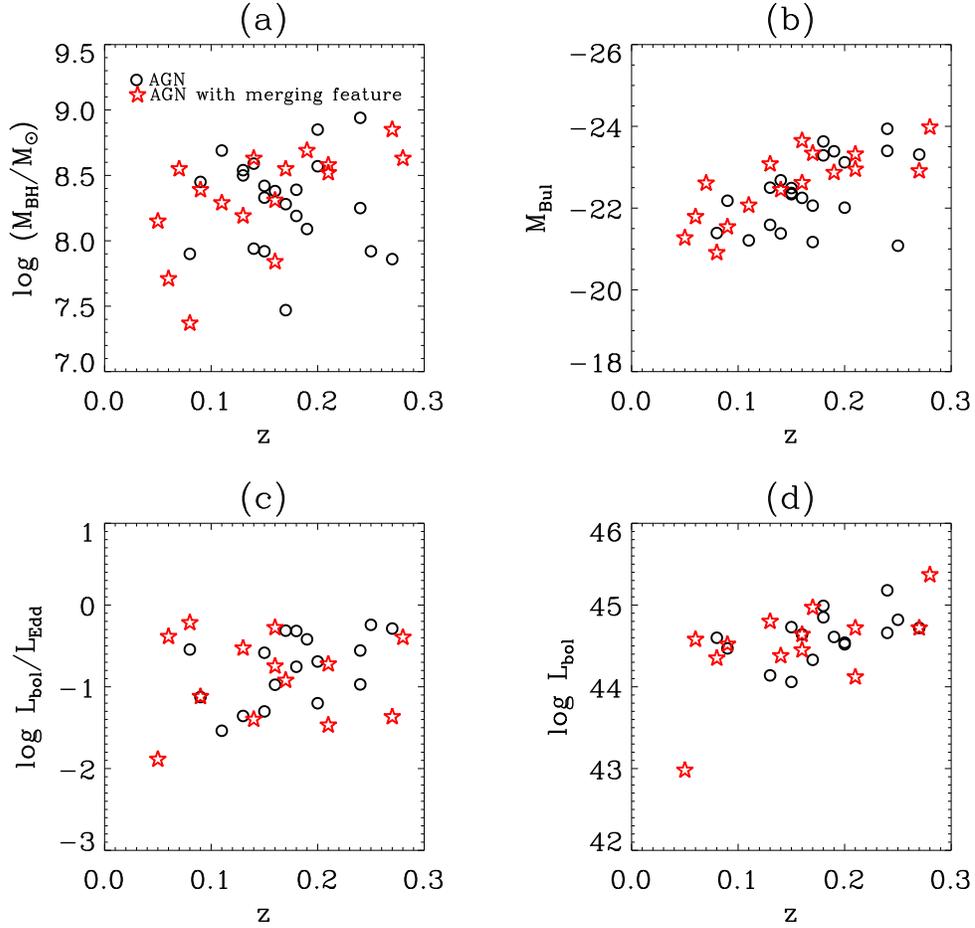}
\caption{Redshift versus (a) BH masses, (b) the host bulge magnitudes, 
(c) the Eddington ratios, and (d) the bolometric luminosities of the base AGN sample
 (black circles) and AGNs with a merging feature as classified in Section \ref{result} (red stars).}.\label{agn}
\end{figure}

 Figure \ref{agn} shows the redshift versus the BH mass, the Eddington ratio, the bolometric 
 luminosity, and the host bulge magnitude of our sample.
 Our sample spans parameter spaces of $10^{7.3} M_\odot \sim 10^{9} M_\odot$,
  $-2 < \mathrm{log}(L_{\rm bol}/L_{\rm Edd}) < 0$,  $43 < {\rm log}(L_{\mathrm{bol}}/L_{\odot}) < 45.5$,  and -24 mag $< M_{\mathrm{Bul}}(R) < -20.5$mag. 

\section{CONTROL SAMPLE AND SIMULATION OF AGNs\label{sample selection_cntl}}

\subsection{Control Sample}\label{control}

 We look into the merging fraction of AGNs through the presence of a merging feature in AGN host galaxies.
 To compare it with the merging fraction of inactive galaxies, we construct a control sample. 
 The control sample is restricted to only early-type galaxies because the majority of host galaxies
 in our luminous AGN sample are early-type galaxies. For the control sample, 
 we chose the SDSS Stripe 82 early-type galaxies
 with a surface-brightness limit of ${\mu}_r$ = 27 mag arcsec$^{-2}$ \citep{kaviraj},  
 which is comparable to or deeper than the image depths of our AGN images \citep{kaviraj} when taking into account the surface-brightness dimming effect in the images of the AGN sample. 
 There are 317 early-type galaxies in this sample, which is limited to be at $z < 0.05$ and $M_{r} < -20.5$mag. Table \ref{tbl-control} shows the basic properties of the control sample, such as the 
 morphological type, $B/T$, and $M_{\rm BH}$.

 These early-type galaxies have been classified through careful visual 
 inspection by \cite{kaviraj}. The classifications include ``relaxed," ``tidal feature," ``dust feature,"  ``tidal and dust feature," ``interacting," and 
 ``spheroidal galaxy with faint disk features." 
 Among these classifications, we adopt ``tidal feature," ``tidal \& dust feature," and ``interacting
 types" as signs for a merging feature. 
  There are 57 early-type galaxies identified as objects with
  merging features.
 A dust feature may be a sign of past merging, but we do not include it as a ``merging feature'' in
 our analysis because dust lanes do not extend to the outer region of galaxies and will be 
 extremely difficult to detect in AGNs because a bright point source outshines the dust feature.
 Note that we simulate AGNs at different redshifts using the control sample in Section 3.2, 
in order to account for biases arising from a redshift difference between the
control and the AGN samples, such as the surface brightness dimming and the change in the
sizes of objects. A potential dependence of the merging fraction as a function of
luminosity will be removed by analyzing the AGN sample at different host galaxy $M_{\rm BH}$
and luminosity bins and taking into account the difference in the $M_{\rm BH}$ distribution
(Sections 3.2 and 4 and Figure 9).

 The median redshift of the Stripe 82 early-type galaxies ($z_{med} = 0.04$) is 
 lower than that of the AGN sample ($z_{med} = 0.15$). An evolution in the merging
fraction could affect the analysis of the merging fraction difference between the control
and the AGN samples, but such an effect is negligible 
($< 20\, \%$), as described below. Earlier works on the study of the merger fraction
were controversial, with some studies showing a strong evolution in redshift that goes with 
$\sim (1+z)^{4}$ (e.g., Zepf \& Koo 1989; Carlberg et al. 1994; Yee \& Ellingson 1995), and with others
indicating a weak evolution ($\sim (1+z)^{1.2}$; e.g., Neuschaefer et al. 1997). However,
 more recent results are converging to a rather weak redshift evolution in 
 the merging fraction of $(1+z)^{0 - 2}$ at $z < 1$ (Lin et al. 2004; Lotz et al. 2008a; Bertone et al. 2009; Bundy et al. 2009; Jogee et al. 2009; Lopez-Sanjuan et al. 2009; Man et al. 2012; Xu et al. 2012), which is especially true for massive or luminous galaxies that are comparable in the mass/luminosity range of the host galaxies studied here 
 (Hsieh et al. 2008; Robaina et al. 2010; ). Even if we adopt the most extreme merging
 fraction evolution of $(1+z)^{2}$ from the recent results, the expected increase in the merging fraction from $z=0.04$ to $z=0.15$ is only at the level of 20\%, which is much 
smaller than the discrepancy in the merging fraction that we are exploring in this work.

\begin{figure}[p]
\includegraphics[width=16cm]{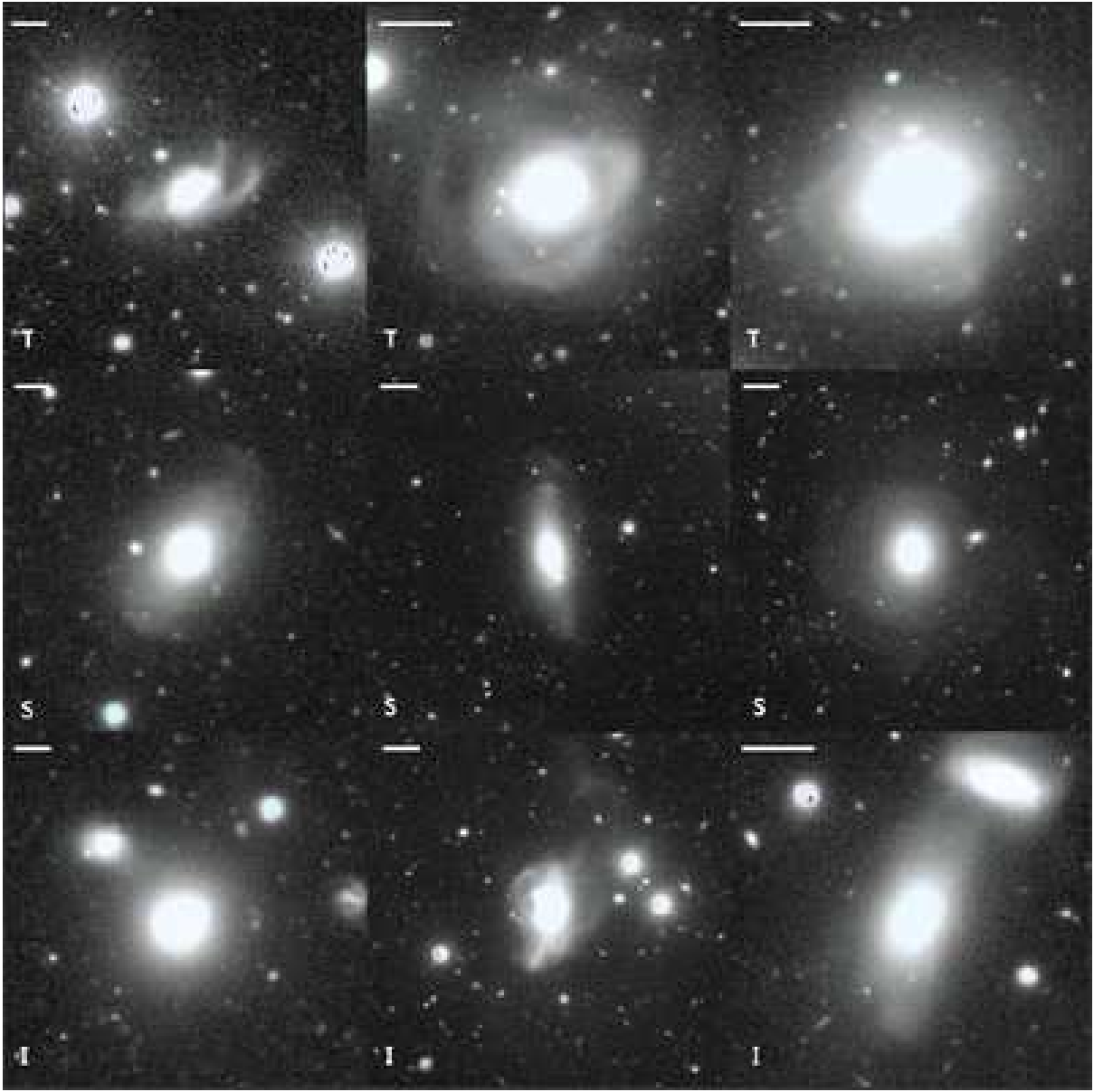}
\caption{Examples of the Stripe 82 early-type galaxies with merging features
classified as tidal tail (T), shell (S), interaction (I) types, respectively.
The length of the horizontal bar in each panel is 10$\arcsec$.
}\label{mergingtype}
\end{figure}

  The merging features are further categorized by us into tidal tail (T), shell (S), or interaction (I), as 
 shown in Figure \ref{mergingtype}. Our definition of ``tidal tail'' includes tidal tails and disturbed features like loops. 
 The ``shell'' types are distinguished from ``T'' types  as objects with distorted surface brightness and shell structures, but lacking features like loops. The ``interaction" types are objects with their surface-brightness profile extending to a close companion. If there is no merging feature (i.e., ``relaxed type'' in
 Kaviraj 2010), we categorize such a case as no feature (N).
  Although there are quantitative and automatic classification techniques, 
we adopted the classification through visual inspection  
 because the automatic classification may miss some faint features \citep{Lotz, Kartaltepe, adams2012}.

\begin{figure}[p]
\includegraphics[width=16cm]{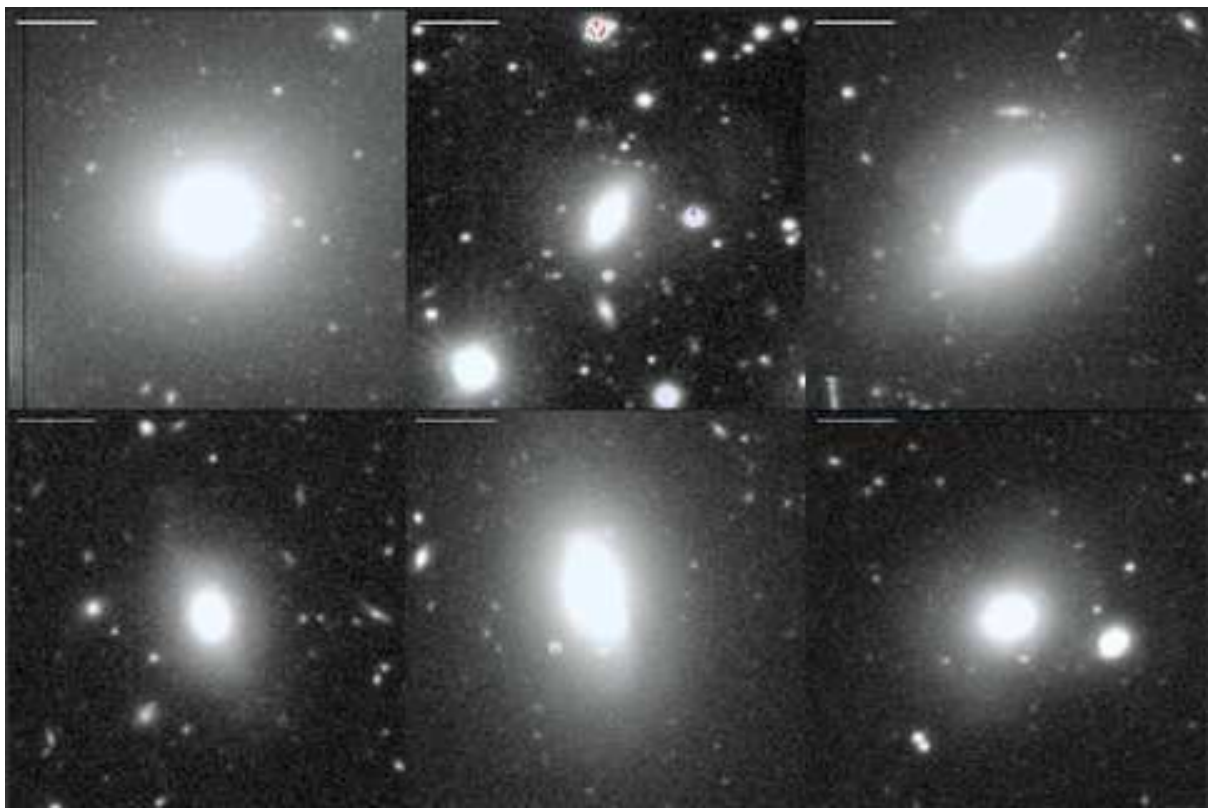}
\caption{Examples of galaxies with ambiguous classification.
They are classified as merging systems by \cite{kaviraj}. However, we do not identify the merging feature,
and it is likely that the merging feature, even if one assumes that it exists,
will not be identified as such by us in our AGN sample. The length of the horizontal bar in each panel is 10$\arcsec$.
}\label{ambiguous}
\end{figure}

 During the visual reinspection of the SDSS Stripe 82 images, we found six galaxies 
 classified as objects with a merging feature in \cite{kaviraj}, but we felt that it is not clear 
 whether they really have merging features or not (Figure \ref{ambiguous}). We treat them 
 in our analysis as early-type galaxies with a merging feature, although we are unlikely 
 to classify AGNs
 with this kind of object as AGNs with a merging feature.

 The bulge magnitudes of the control sample early-type galaxies are taken from the catalog of \cite{simard}, who performed a bulge+disk
 decomposition of galaxies in the SDSS DR7. 
 We adopted the fitted parameters from a model with free Sersic index $n$ for the bulge component and $n=1$ for the disk component.
 This catalog covers about 84\% of the control sample. \cite{simard} did not perform the bulge+disk fitting for the remaining 
 16\% of galaxies because they are brighter than $m_{\rm petro}(r) < 14$ mag (too large to fit with within their automated pipeline). 
 For these bright early-type galaxies, we assume that they are bulge-dominated galaxies.
 Note that the SDSS $r$-band magnitude is converted into the R-band magnitude ($r-R = 0.151$ mag), assuming an elliptical galaxy spectrum of 
 \cite{calzetti}. The black hole masses were estimated using Eq. (2), but with the $\alpha=-2.6$ and $\beta=-0.5$ of inactive
 galaxies \citep{bettoni}.

\begin{figure}[p]
\includegraphics[width=16cm]{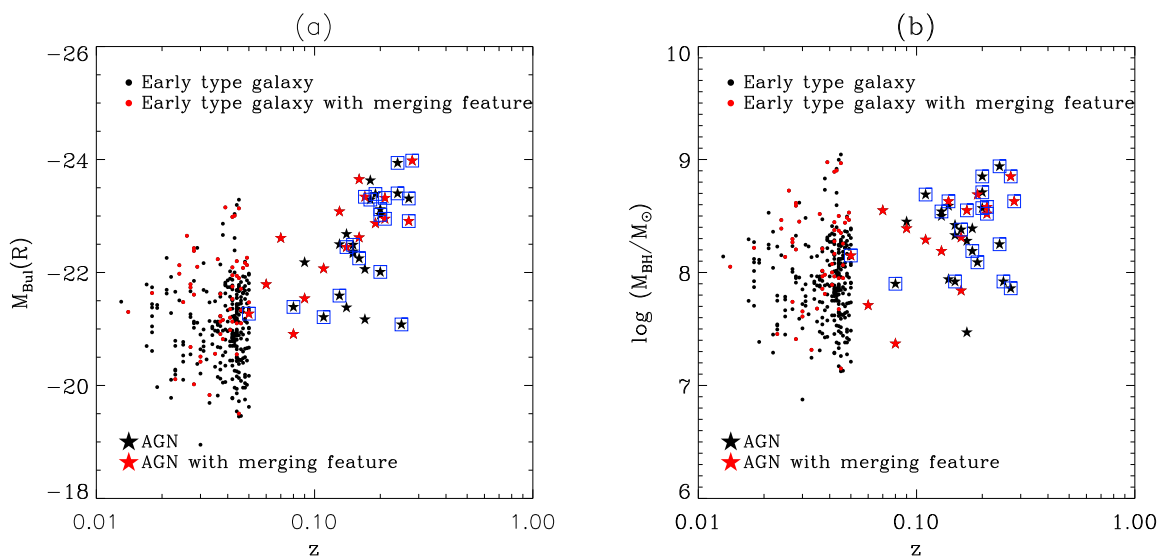}
\caption{(a) Bulge magnitudes in the $R$ band of the control sample early-type galaxies (small black circles), 
 the control sample early-type galaxies with a merging feature (small red circles),
 AGN hosts (black stars), and AGN hosts with a merging feature (red stars) as a function of redshift, 
(b) Black hole masses of the control sample early-type galaxies (black circles), 
the control sample early-type galaxies with a merging feature (red circles),
 AGN hosts without (black stars) and with (red stars) merging feature as a function of redshift. The blue rectangles denote AGNs from the DuPont sample.}\label{magc}
\end{figure}

\begin{figure}[p]
\includegraphics[width=16cm]{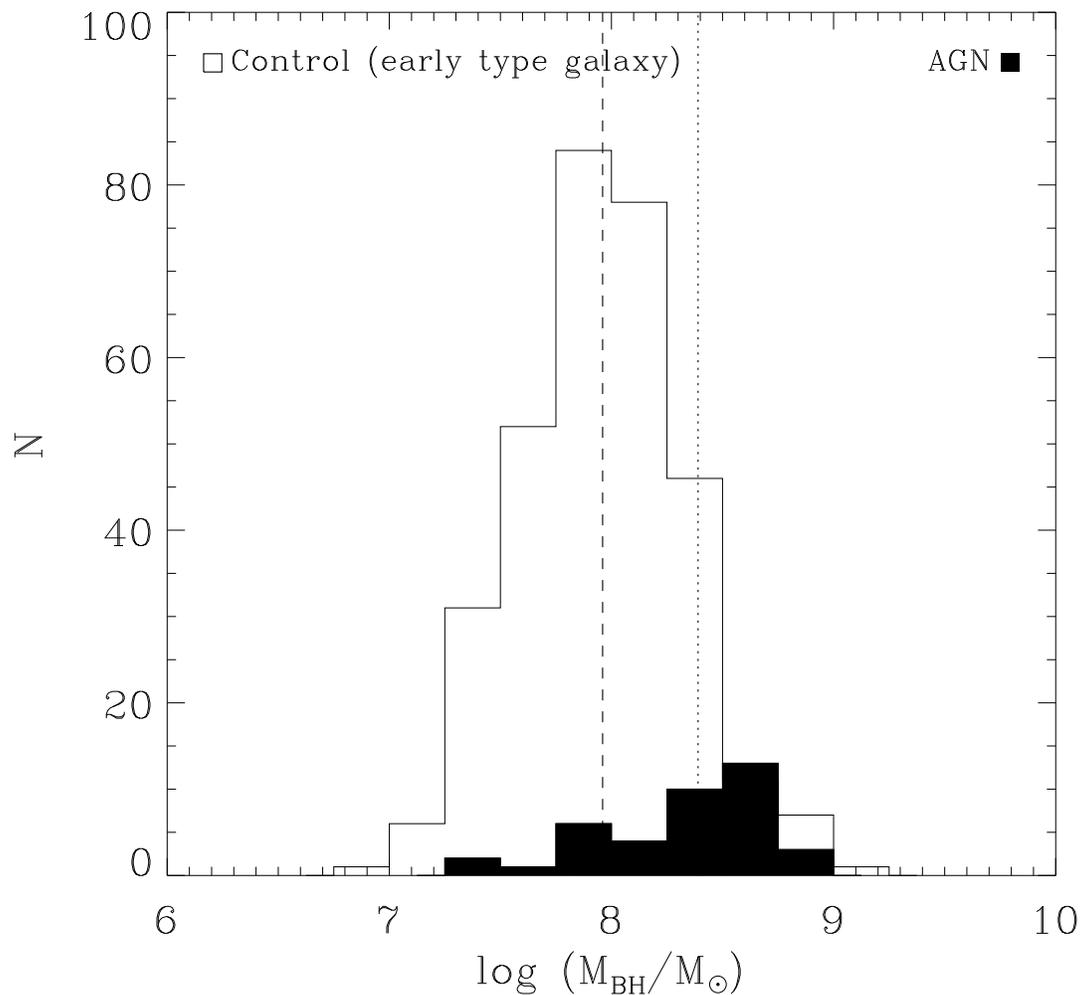}
\caption{Black hole mass distribution of the control sample early-type galaxies and AGNs. 
The unfilled black histogram and the dashed line are the black hole mass distribution and its median ($10^{7.96} M_\odot$)
of the control sample early-type galaxies. The black filled histogram and the dotted line are
the black hole mass distribution and its median ($10^{8.39} M_\odot$) of the AGNs.}\label{bhc}
\end{figure}

\begin{figure}[p]
\includegraphics[width=16cm]{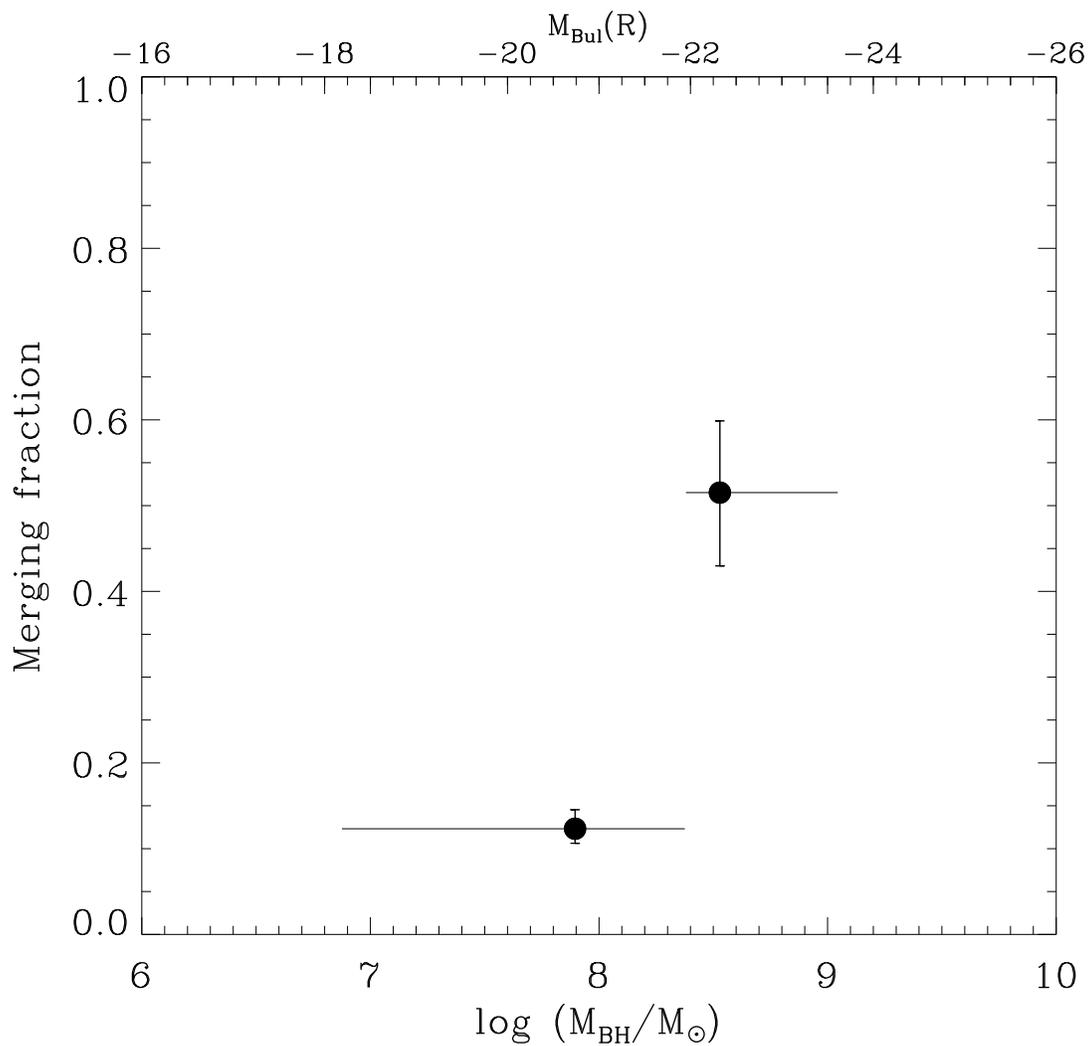}
\caption{Merging fraction of the control sample early-type galaxies 
as a function of $M_{\mathrm{BH}}$ and $M_{\mathrm{Bul}}$.
 The merging fraction increases as $M_{\mathrm{BH}}$ and $M_{\mathrm{Bul}}$ increases.
 The horizontal bars show the $M_{\mathrm{BH}}$ and $M_{\mathrm{Bul}}$ ranges covered by early-type galaxies
in two bins divided at $10^{8.39} M_\odot$ and -21.81 mag.}\label{mf}
\end{figure}

 Figure \ref{magc} shows the distribution of the bulge absolute magnitudes in $R$ and the black hole masses of the control sample as a function of $z$. 
 Overplotted in the figures is our AGN sample.
 Figure \ref{bhc} compares the $M_{\rm BH}$ distribution of the control sample and that of the AGN sample. 
 There is a difference in the overall parameter space covered by the AGN and the control
 samples, where the
 $M_{\mathrm{BH}}$ values of the AGN base sample tend to be brighter and more massive than for the control sample. 
 This difference can introduce a bias in the comparison of merging fractions 
 if the merging fraction changes 
 as a function of host galaxy brightness or black hole mass. Figure \ref{mf} shows
 the merging fraction as a function
 of $M_{\mathrm{BH}}$ and $M_{\mathrm{Bul}}$ of early-type galaxies in the control sample.
  The error bars indicate 68.3\% confidence levels, which are computed using the method
  described in Cameron (2011)\footnote{We will use this method to compute errors of fractional
  quantities throughout the paper.}.   
   We find that the merging fraction increases with $M_{\mathrm{BH}}$ or the bulge luminosity. 
 Therefore, we will take into account the difference in the parameter space 
 ($M_{\mathrm{BH}}$ or $M_{\mathrm{Bul}}$) between the AGN and the control samples in our analysis, and we will use $M_{\mathrm{BH}}$ as the reference parameter
 to adjust for the difference in the covered parameter space. We adopt $M_{\mathrm{BH}}$ instead of $M_{\mathrm{Bul}}(R)$,
 because it is a better defined quantity for the AGN sample, and the $M_{\rm BH}$ versus $M_{\rm Bul}$ relation is established more securely 
for passive galaxies (the control sample) than for AGNs.

\subsection{Simulation of AGNs Using a Control Sample\label{simulation_control}}

 It is well known that 
 most (80\%) of early-type galaxies show merging features in very deep images 
\citep[$\mu \approx$ 28 mag arcsec$^{-2}$;][]{vanDokkum}, so a careful analysis
 is needed to analyze AGNs that are associated with merging. 
The fraction of early-type galaxies with a merging feature is 
 17\% for the control sample \citep{kaviraj}, but there are many factors that can influence 
 the detection rate of a merging feature in the AGN sample.

  First, the depth of the image is important because many early-type galaxies with merging features 
 do not reveal merging features in a shallower image. 
 We tried to match the depth of our AGN images and the control sample image, 
 but not perfectly. Most notably, the rest-frame (effective) surface-bright limit of the AGN sample is a bit shallower 
 by 0.5 - 1 mag arcsec$^{-2}$ at $z=0.2$-$0.3$ than that of the control sample ($z_{\rm med} \sim 0.04$) due to the surface brightness dimming.

 Second, the AGNs are located farther away than the control sample,
 so their apparent sizes are smaller. This causes a loss
 in the resolution element, which can wash away some of the merging features. 
 This resolution effect also needs to be treated properly.

  Third, AGNs have bright nuclear components. For AGNs at a very low redshift, 
 the PSF widths are small compared to the extent of the host galaxy, so the 
 the bright nuclear component does not hamper the detection of a merging feature. 
  However, if they are located at a higher redshift,
  the width of the PSF becomes comparable to or larger than the host galaxy size,
 and when combined with an extended wing of the PSF in many ground-based images,
 low surface-brightness features in the host galaxy becomes difficult to detect.

 In order to properly handle these biases, 
 we perform a simulation where we take the control sample early-type galaxy images, 
plant a nuclear component, and place them at various redshifts.
 The simulation is done in the following way.

 In the first step, early-type galaxy images are block averaged by a factor of 2--5, 
 which corresponds to the change in  the angular diameter distance $D_{A}(z)$ 
 from $z=0.041$ (the median redshift of the control sample) to the redshifts
 of 0.086, 0.138, 0.2, and 0.275, respectively, matching the redshifts of the AGN sample.
  The pixel scales of the block-averaged images are all considered as 
 $0\farcs3961$/pixel, as in the original Stripe 82 image, to properly reflect the redshift changes.

   Because the simulated images are block averaged, 
  FWHM of the PSF given in pixel units has shrunk. To mimic the  
  PSF resolution (in arcseconds) of the AGN images, 
 we convolved the block-averaged images using a Gaussian function with a $\sigma_{G}$ value 
 as \[ \sigma_{\rm G}^{2} = \sigma_{\rm out}^{2} - (\frac{\sigma_{\rm S82}}{f})^{2}, \] where
 $\sigma_{\rm S82}$ is FWHM/2.35 of the image before block averaging, $f$ is the block-averaging factor, and
 $\sigma_{\rm out}$ is the desired resolution of the output image after the convolution. We adopt two values of
 $\sigma_{\rm out}$ so that the seeing of the resultant image becomes 
 either 1$\arcsec$ or 2$\arcsec $ 
 in FWHM to allow us to investigate 
 the effect of seeing in the identification of a merging feature.

 The pixel values are adjusted so that the total flux of the simulated galaxy is equal to the 
 flux of the galaxy at the desired redshift.
 During the block-averaging process, the background noise is smoothed out.
 To make the surface-brightness limit of the resultant image to be $\sim 27$ mag arcsec$^{-2}$,
 additional noises are implemented in the simulated images.

 Finally, we plant a point source at the center of the simulated early-type galaxy images 
in order to mimic an AGN and investigate 
 the effect that a bright AGN component has on the merging fraction analysis. 
 We find that the median of the
 differences between the host magnitude, $m_{\mathrm{host}}$, and the nuclear magnitude, $m_{\mathrm{Nuc}}$, 
is $(m_{\mathrm{host}} - m_{\mathrm{Nuc}})_{\mathrm{med}} =  \sim 1.0$ mag, 
 and the upper and the lower quartiles of the $m_{\mathrm{host}} - m_{\mathrm{Nuc}}$ values 
 from the mode of the $m_{\mathrm{host}} - m_{\mathrm{Nuc}}$ distribution are $\sim$ 1.5, 0 mag. 
 To simulate the nuclear point source, we chose a bright star in the SDSS Stripe 82 image
 and rescaled its pixel values so that the PSF has three representative values of
 $m_{\mathrm{gal}} - m_{\mathrm{Nuc}}$ that are 0, 1.0, and 1.5 mags 
brighter than the host magnitude.
 These PSFs are put at the center of the simulated early-type galaxy images.

\begin{figure}[p]
\includegraphics[width=16cm]{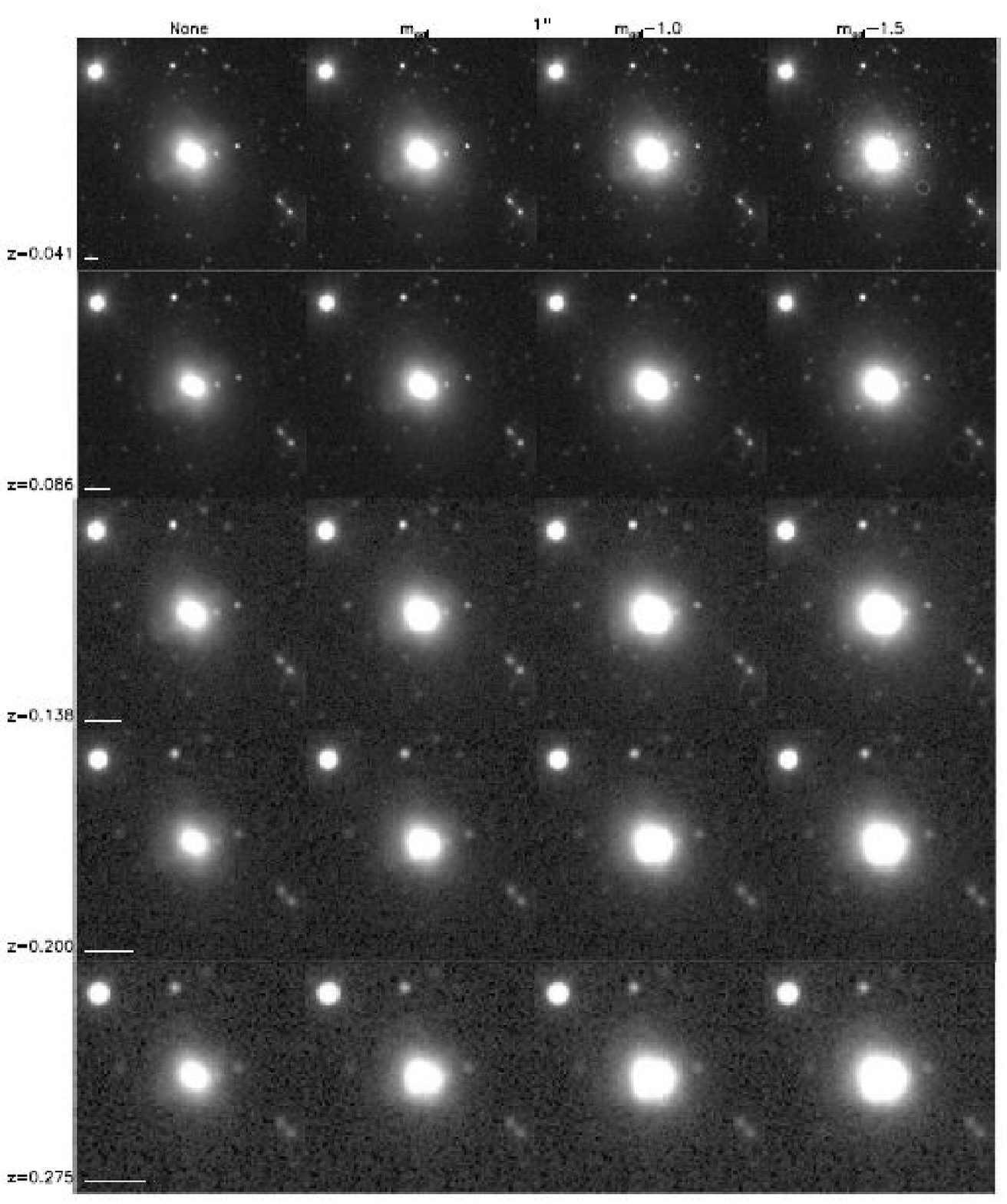}
\caption{Examples of simulated AGN images. The seeing is 1$\arcsec$. The redshifts and the magnitudes of
 the nuclear point sources are varied as indicated (see text).
The length of the horizontal bar in each panel is 10$\arcsec$.} \label{simul1}
\end{figure}

\begin{figure}[p]
\includegraphics[width=16cm]{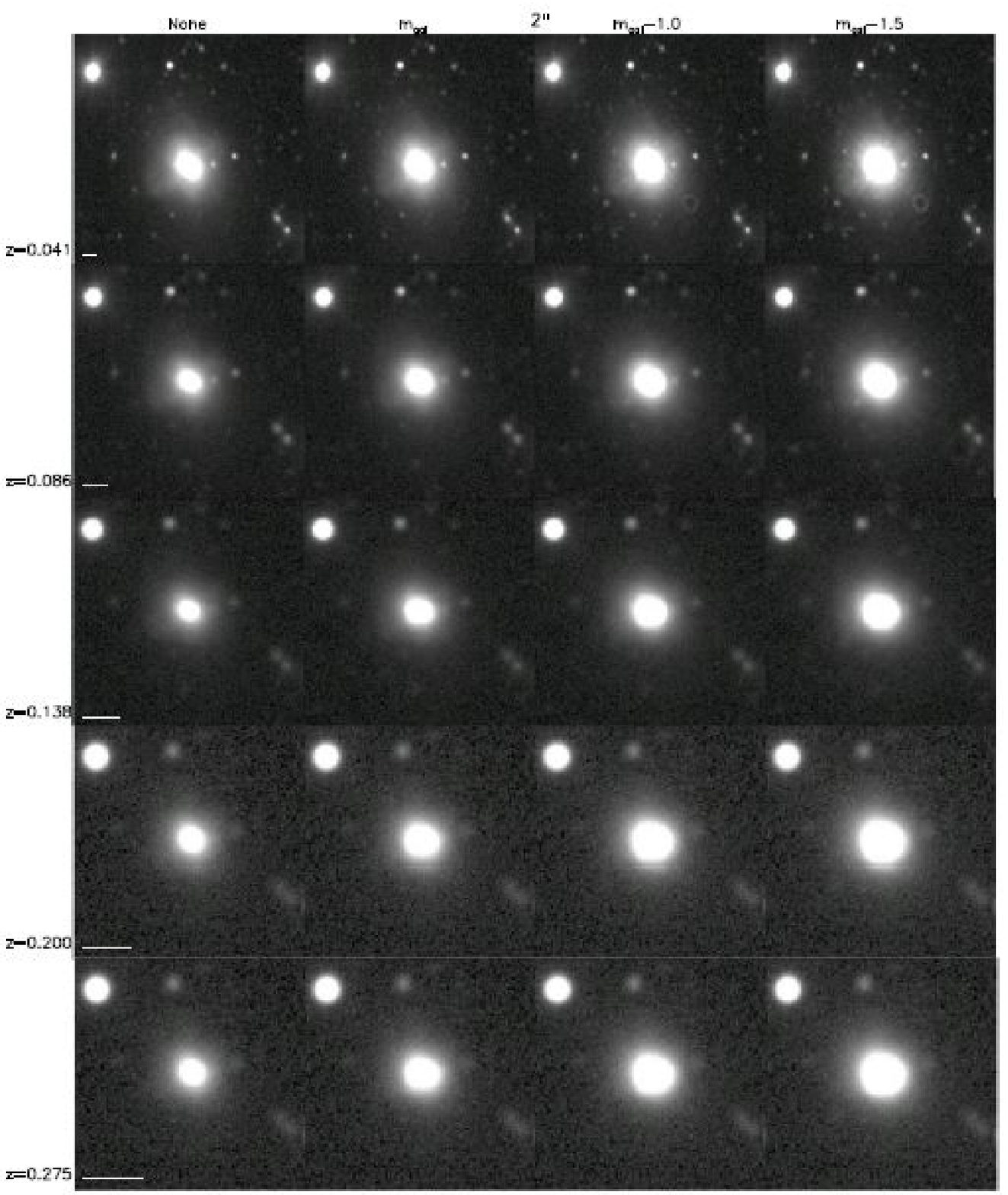}
\caption{Examples of simulated AGN images. The seeing is $2\arcsec$. The redshifts and the magnitudes of
 the nuclear point sources are varied as indicated (see text).
The length of the horizontal bar in each panel is 10$\arcsec$.} \label{simul2}
\end{figure}

\begin{figure}[p]
\includegraphics[width=16cm]{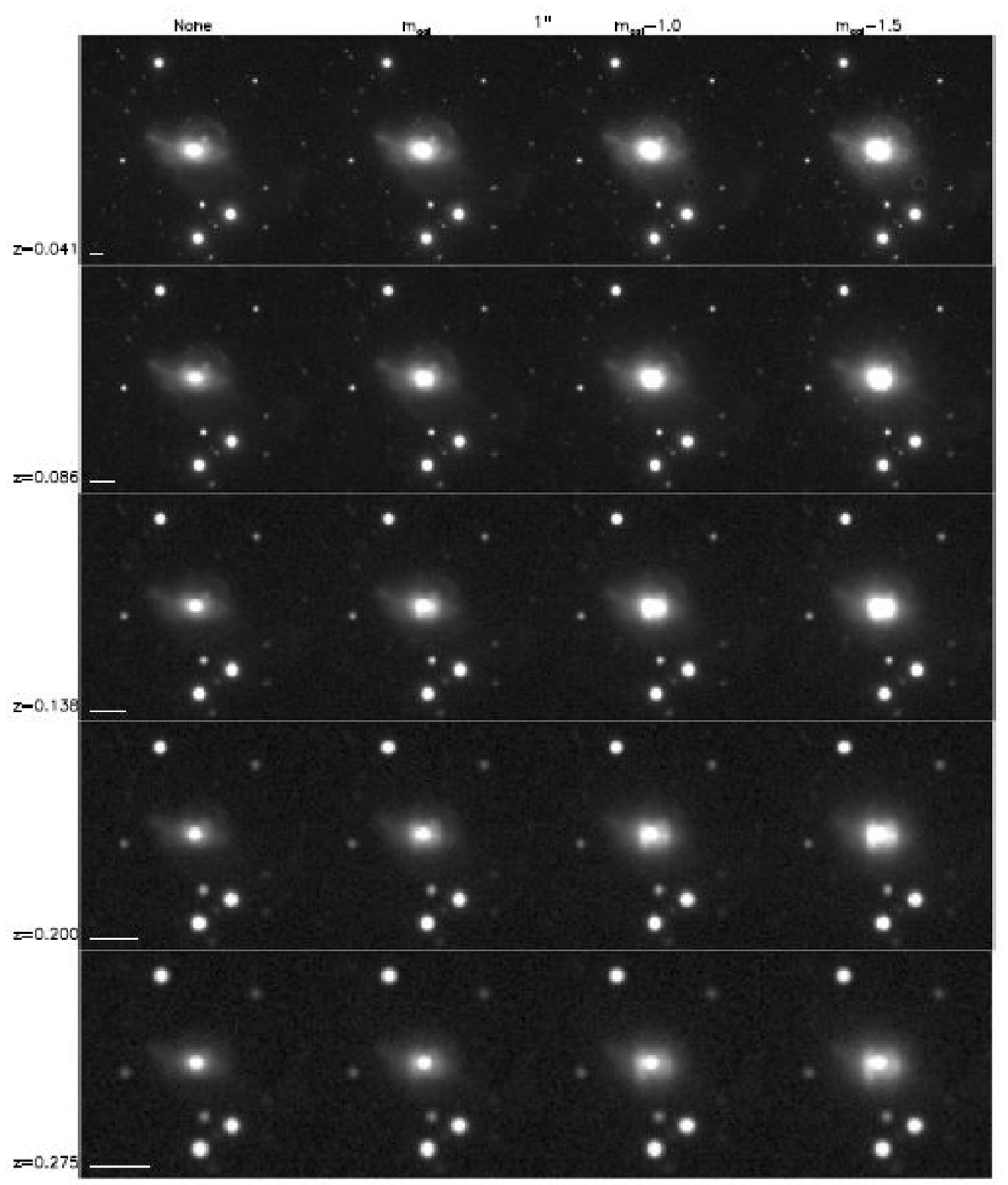}
\caption{Examples of simulated AGN images. The seeing is 1$\arcsec$. The redshifts and the magnitudes of
 the nuclear point sources are varied as indicated (see text).
The length of the horizontal bar in each panel is 10$\arcsec$.} \label{simul3}
\end{figure}

\begin{figure}[p]
\includegraphics[width=16cm]{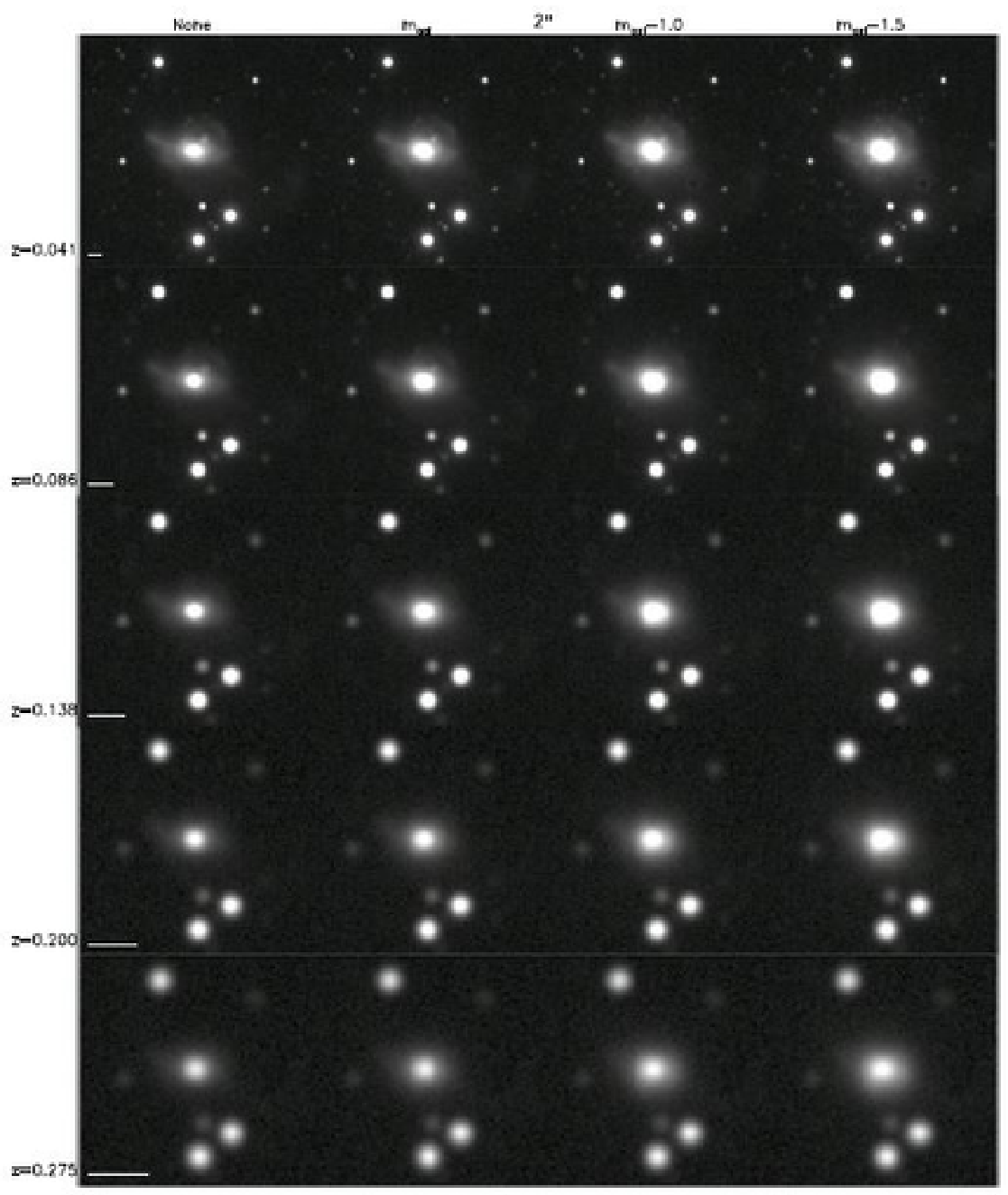}
\caption{Examples of simulated AGN images. The seeing is $2\arcsec$. The redshifts and the magnitudes of
 the nuclear point sources are varied as indicated (see text).
The length of the horizontal bar in each panel is 10$\arcsec$.} \label{simul4}
\end{figure}

 In summary, we made simulated AGN images for each of 57 SDSS Stripe 82 early-type galaxies
 with merging features,
 in a parameter that covers $z=$ (0.041, 0.086, 0.138, 0.2, 0.275), $m_{\mathrm{Nuc}}=$ (none, 
 $m_{\mathrm{gal}}$, $m_{\mathrm{gal}} - 1$, $m_{\mathrm{gal}}-1.5$), and seeing= ($1\arcsec$, $2\arcsec$), leading to
 the creation of 40 sets of simulated AGN images per a SDSS Stripe 82 early-type galaxy. 
Figures \ref{simul1}, \ref{simul2}, \ref{simul3}, \ref{simul4} show simulated AGN images based on three SDSS Stripe 82 early-type galaxies. 
Figures \ref{simul1}, \ref{simul3} are for the seeing of 1$\arcsec$, and Figures \ref{simul2}, \ref{simul4} are for the seeing of 2$\arcsec$.
Redshifts are 0.041, 0.086, 0.138, 0.2, and 0.275 from top to bottom, and the 
magnitudes of the nuclear point sources are none, $m_{\mathrm{gal}}$, $m_{\mathrm{gal}}$-1.0, and $m_{\mathrm{gal}}$-1.5 mags
 from left to right.

\begin{figure}[p]
\includegraphics[width=14cm]{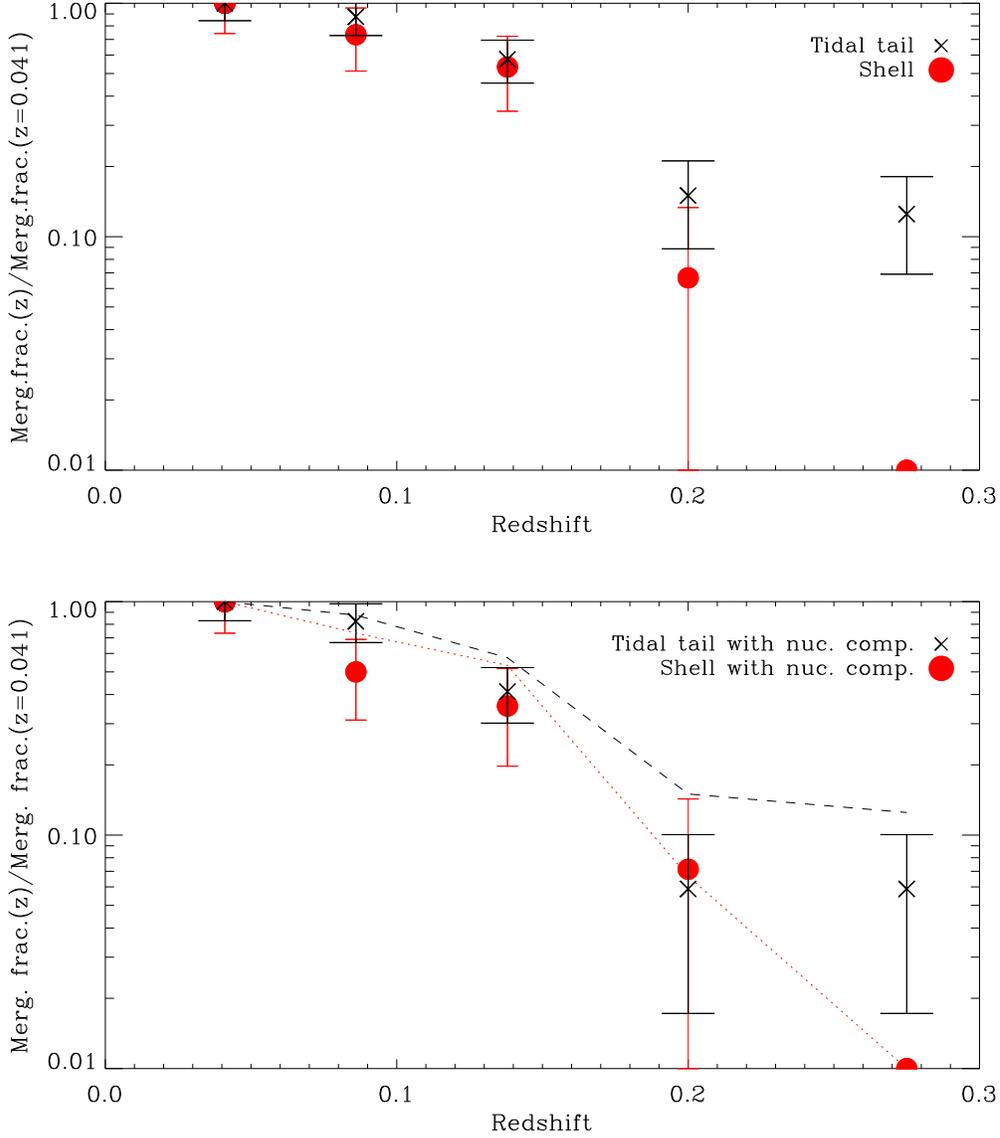}
\caption{Change in the merging fraction when simulated AGNs are redshifted from $z=0.041$ to a higher redshift but without adding a nuclear component (the upper panel) and when the simulated AGNs are redshifted and a nuclear component is added (the bottom panel). The magnitude of the nuclear component is assumed to be $m_{\mathrm{Nuc}}=m_{\mathrm{host}} - 1.0$ mag. Cases for two types of merging features (black cross for T type and red filled circles for S type) are indicated. The dashed and the dotted lines in the bottom figure denote the result for the T type and the S type in the upper figure, respectively.}\label{mf_ts}
\end{figure}

 Figure \ref{mf_ts} shows quantitatively how the merging features like tidal tails and shells 
become difficult to identify when the redshifts are changed and the nuclear sources are added.  
 The top panel is for the case that does not include the nuclear point source, and it shows how the redshift effect alone 
affects the identification of a merging feature.
 The bottom panel in the figure assumes the addition of a nuclear source with $m_{\mathrm{gal}} - 1$ mag 
and a seeing of $1\arcsec$ and shows how the nuclear source makes it more difficult
to see the merging features. We find that there is a steady decrease in the identification 
of a merging feature as we move objects to higher redshifts. With respect to $z=0.041$, only 50\% of the features can be identified at $z=0.14$, and this value goes down to 5 -- 10\% at $z=0.2$.
 The shell features are more difficult to identify than the tidal features, and 
all of the shell features are virtually unidentifiable at $z > 0.2$. 
 If we include a bright point source, the identification rate of merging features drops by 10 - 20\% (or by a factor of two at an intermediate redshift). 
Our simulation demonstrates that the merging features can disappear easily
 at a higher redshift and with an addition of a bright nuclear component. Hence, one must take
 these effects into consideration when comparing the merging 
 fraction of AGNs and with of the control sample galaxies.

\section{RESULTS}\label{result}

\subsection{Merging Features in the AGN Sample\label{visual_agn}}

  Figure \ref{merging1} shows the images of our AGN sample in the order as listed in Table
 \ref{tbl-1}. Using Figure \ref{merging1}, we identified the merging features through visual inspection.
  The merging features are categorized into  tidal tail (T), shell (S), or interaction (I), and
 no feature (N), as was done for the control sample.
  The classification of the merging features is indicated in Table \ref{tbl-1} and also in Figure \ref{merging1}

  Our visual inspection reveals that 17 of 39 AGNs (43.6\%) show a merging feature. Among these AGNs with a merging feature,
 only one AGN shows a shell structure, and another shows a sign of interaction. Therefore, the majority of AGNs showing
 merging features are categorized as T.

\subsection{Merging Fraction: Comparison with Simulated AGNs}\label{result_simul}

 In the previous section, we find that 17 of 39 AGNs ($\sim$ 43.6\%) show merging features, but
 only $\sim$17\% of the control sample shows evidence of galaxy mergers.
 At face value, the merging fraction of AGNs is $\sim$2.6 times higher than the merging fraction of the control sample.
 However, the difference in the merging fraction should be more pronounced if we consider other effects that could make
 it difficult to identify merging features in the AGN sample.
 For example, Figures \ref{simul1} - \ref{simul4} indicate that it becomes very difficult to identify merging features when
 an object is at a higher redshift and has a brighter nuclear component. We will take these effects into account
 and derive a more realistic comparison of the merging fraction between the control and the AGN samples.
  On the other hand, there are several AGN images that are deeper than the
 control sample images (the points above the dotted line in Figure 3). For these
 cases, the noises are added to the AGN images to match the control sample image 
 depth, and the merging feature is classified on the noise-added images. We find 
 that the merging features can be identified in the noise added
 images, which is not surprising considering that the depths of these AGN images
 are only a few tenths of magnitude deeper.

   Because the merging fraction changes as a function of
 $M_{\rm BH}$ (or equivalently, $M_{\rm Bul}(R)$), as was shown for the control sample in Figure 10,  
  it is important that we use the control galaxies that have the same properties as the galaxies that AGN hosts evolve into. One could use $M_{\rm Bul}(R)$ to tag control 
 galaxies equivalent to an AGN host, but this is not desirable because the growth of AGN hosts has
 been found to be significant and the exact amount of the evolution is uncertain 
\citep[e.g.,][]{woo06, woo08, hiner2012}.
Moreover, the measurement of $M_{Bul}(R)$ is challenging for AGNs with a luminous nuclear component
 (Sanchez et al. 2004; Simmons et al. 2008;
 Kim et al. 2008b). Using the $M_{\rm BH}$ of an AGN host to identify its counterpart 
 in the control sample is 
 advantageous in this regard because BHs seem to have already grown at $z \sim 0.3$ in active galaxies, and $M_{\rm BH}$ estimators are well established for both AGNs and quiescent galaxies. 
 However, the use of $M_{\rm BH}$ is not completely free from difficulties: the intrinsic scatter can be 
 as large as 0.4 dex for $M_{\rm BH}$ estimators (e.g., \citep{Peterson2004}), and BHs could still grow 
 in high $L_{\rm bol}/L_{\rm Edd}$ AGNs (Kim et al. 2008). 
  For the following analysis, we consider $M_{\rm BH}$ as a more desirable quantity to match
 AGNs with equivalent galaxies in the control sample, but we also present results that use    
 $M_{\rm Bul}(R)$ as a galaxy identifier in order to check if the choice of the galaxy identifier 
 can change the main conclusion.

    Figure \ref{mf_z} shows the merging fraction of the simulated AGNs as a function of redshift. 
  We divide the sample into two, one with
 $\mathrm{log} (M_{\rm BH}/M_{\odot}) \geq 8.39$ and another with $\mathrm{log} (M_{\rm BH}/M_{\odot}) < 8.39$, 
 because the merging fraction changes as a function
 of $M_{\mathrm{BH}}$ as we have shown earlier (Section \ref{control}). 
  The panels for each $M_{\mathrm{BH}}$ are also divided into three, 
 each corresponding to a different nuclear magnitude.
  The magnitudes of the PSF components for each panel are $m_{\mathrm{gal}}$ (top), 
 $m_{\mathrm{gal}}-1.0$ (middle), and $m_{\mathrm{gal}}-1.5$ (bottom). 
  The black asterisks are the merging fraction of early-type galaxies without any nuclear
component included, and
 the black open circles are for the cases with a nuclear component included. 
 The seeing of the simulated images is assumed to be 1$\arcsec$ here.
  Tables \ref{tbl-3}, \ref{tbl-4} summarize the merging fractions in the figure, 
 as well as similar numbers for the case of seeing at $\sim 2 \arcsec$.

  Figure \ref{mf_z} shows that the merging fraction of the simulated AGNs decreases 
as a function of redshift,
 even without adding a nuclear component. At $z=0.041$, it is 17\%, 
but it decreases to 10\% at $z=0.138$ (a drop of
 40\%) and to 2\% at $z=0.275$. The effect of the addition of  
 the nuclear component becomes prominent when the nuclear component
 is $m_{\mathrm{Nuc}} = m_{\mathrm{gal}} - 1$ mag or brighter and at $z > 0.138$. At $z \geq 0.2$, 
the merging fraction drops by 50\% or more due to the 
 addition of the nuclear component.
  A larger seeing makes it more difficult to identify merging features. At $z < 0.2$, 
 the effect of the seeing is not much, with a decrease in the merging fraction by 20--30\%. 
 But at $z \geq 0.2$, it could decrease the merging fraction by 50\% or more.

  In Figure \ref{mf_z}, we plot the merging fraction of our AGN sample too (the red stars). 
  The AGN sample is divided into two redshift bins at $z=0.155$ 
 to provide enough number statistics in each bin. 
  The horizontal bars on each point show the redshift range covered by the AGNs in each bin. 
  The locations of the red points show the median redshift in each bin. We find that about 40 - 50\% of AGNs with 
 $M_{\mathrm{BH}} \geq 10^{8.39}\, M_{\odot}$ have merging features regardless of redshift. On the other hand, the merging fraction 
 of AGNs with $M_{\mathrm{BH}} < 10^{8.39} \, M_{\odot}$ varies from 60\% to 20\% going from low to high redshift.
  Note that the actual merging fraction of AGNs could be higher because the redshift shift and 
 the bright nuclear component can make the merging feature difficult to identify, as we saw from the simulation in
 the previous section.  
  In comparison to the simulated AGNs, the merging fraction of AGNs is much higher by a factor of 3--10, 
 except for the redshift bin at $z=0.12$ of AGNs with massive black holes.

  The result is combined into Figure \ref{mf_bh_sum}, where we plot the merging fraction of
 AGNs at two different black hole mass bins,
 and compare it to the same quantity derived for the simulated AGNs (the seeing value 
 of 1$\arcsec$ is assumed). 
  The red stars show the merging fraction of AGNs in the base
 sample in two mass bins 
 ($7.3 <$ log$({M_{\rm BH}/M_\odot}) < 8.39$ and $8.39 \leq$ log$({M_{\rm BH}/M_\odot}) < 9.0$).
  The red filled stars show the merging fraction of AGNs in the best sample. 
  The black filled circles are the merging fraction of the simulated AGNs with 
$m_{\mathrm{Nuc}} = m_{\mathrm{gal}} - 1$ mag, i.e.,
 the median of the nuclear magnitudes of the AGN sample, and the black points are plotted 
 at the median $M_{\mathrm{BH}}$ of each $M_{\mathrm{BH}}$ bin. 
When calculating the merging fraction of the simulated AGNs, we use Eq.(\ref{eq:fmerg_sim}) 
 in order to take into account   
 the difference in the $M_{\rm BH}$ distribution between the control sample and the AGN sample
 (Figure 9), the spread in the redshift distribution of AGNs (Figure 5):

\begin{equation}\label{eq:fmerg_sim}
\mathrm{Merging~fraction} = \frac{\sum\limits_{i}^{} \sum\limits_{j}^{} f_{corr, \delta M_{\rm BH}(i)}\,f_{corr, \delta z(j)} \,  N_{\rm control,merg}(\delta M_{\rm BH, \it i}, \delta z_{j})}{\sum\limits_{k}^{} f_{corr, \delta M_{\rm BH}(i)}\,N_{\rm control}(\delta M_{\rm BH,\it i})}
\end{equation} 

 Here, the quantity $N_{\rm control, merg}(\delta M_{\rm BH, \it i}, \delta z_{j})$ is the number of  
 simulated AGNs that are identified to have merging features with their $M_{\rm BH}$ and
 redshifts belonging to an $M_{\rm BH}$ bin of $\delta M_{\rm BH, \it i}$ 
 (e.g., $8.25 <$ log$(M_{\rm BH}/M_{\odot}) < 8.5$) and a redshift bin of 
 $\delta z_{j}$, and $N_{\rm control}(\delta M_{\rm BH, \it i})$ is the total number 
 of control galaxies in the mass bin of $\delta M_{\rm BH, \it i}$.   
 The factors $f_{corr, \delta M_{\rm BH}(i)}$ and $f_{corr, \delta z(j)}$ correct for the
 difference in the $M_{\rm BH}$ distribution between the AGN and the control samples 
 and account for a range of redshifts of AGNs.
  They are defined as 
 $f_{corr, \delta M_{\rm BH}(i)} 
 = [N_{\rm control}(\Delta M_{\rm BH})/N_{\rm control}(\delta M_{\rm BH, \it i})]
   \times [N_{\rm AGN}(\delta M_{\rm BH,\it i})/N_{\rm AGN}(\Delta M_{\rm BH})]$
   and  

$f_{corr, \delta z(j)} = N_{\rm AGN}(\delta z_{j}, \Delta M_{\rm BH})/N_{\rm AGN}(\Delta M_{\rm BH})$,
    where $\Delta M_{\rm BH}$ represents either one of the two broad mass bins of 
    $7.3 <$ log$(M_{\rm BH}/M_{\odot}) < 8.39$ and $8.39 \leq$ log$(M_{\rm BH}/M_{\odot}) < 9.0$. 

  Irrespective of whether we adopt the base or the best sample, we find the following (Table \ref{table_mf}). At the higher mass bin of $8.39 \leq \mathrm{log} ({M_{\rm BH}/M_\odot}) < 9.0$, the AGN merging fraction is about $48 \pm 15$\%, and 
 it is $12 \pm 5$\% for the simulated AGNs, which is a difference of a factor of four.
  At the lower mass bin of $7.3< \mathrm{log} ({M_{\rm BH}/M_\odot}) < 8.39$, the AGN fraction is about 39\%, and that of the simulated
 AGNs is only $5\pm2$\%, giving a factor of eight difference. 
 
 Similarly, we plot the merging fraction as a function of the absolute magnitude of the bulge (Figure \ref{mf_mbul_sum}). The same method is adopted for computing the merging fraction as for the $M_{\rm BH}$-based
 merging fractions but using $M_{\rm Bul}(R)$ as a quantity to compute the correction factor. We 
 find exactly the same trend as in Figure \ref{mf_bh_sum}, although the difference in the 
 merging fraction between the control and the AGN samples is reduced in
  at the lower luminosity bin (with respect to the lower $M_{\rm BH}$ bin).
 The reason for the discrepancy can be understood when we examine the difference in the overall 
 distribution of AGNs in $M_{\rm BH}$ versus $M_{R}$
 with respect to the control sample, as seen in Figure \ref{magc}.  
 In $M_{R}$, most AGNs occupy the luminous end of the control sample, while it is less so in
 $M_{\rm BH}$. Luminous early types in the control sample have a higher merging fraction than
 fainter early types, which leads to the higher merging fraction at the lower luminosity bin 
 of the control sample.
  However, the amount of luminosity evolution is uncertain for early-type host galaxies of AGNs, 
  and the amount of luminosity dimming can be well above a few tenths of the magnitude of 
  normal early-type galaxies \citep[e.g.,][]{im96, im2002, bernardi03} 
  from $z=0.3$ to $0.04$.  
  If we make such a correction for the luminosity dimming in the AGN host galaxy,
  the $M_{R}$ distribution would resemble the $M_{\rm BH}$ distribution more closely, 
  and the merging fraction 
  of the control galaxies at the lower luminosity bin would decrease to the level of 
  the value for the lower $M_{\rm BH}$ bin.

  We note that our classification using visual inspection is not likely to affect our result. 
  To quantify 
 possible errors in the visual classification, the visual classification was done by two 
 authors independently for the simulated AGN sample. 
  We find that the classification agrees fairly well, and the resultant error in the 
 merging fraction is only about $\pm$0.01 to $\pm$0.03. Overall, this is a negligible amount. 
 
  We also note that a slight difference in the redshift ranges
 of the control sample and the AGN sample does not bias our result in favor of higher merging fraction in the AGN sample. As we already argued in Section 3.1, the merging fraction evolves 
 only mildly over the redshift range of $z=0.15$ and $z=0.04$. The redshift evolution trend 
 can be reduced by limiting
  our AGN sample to those at $z < 0.1$. Doing so only makes our result stronger, 
  with five out of six AGNs
  at $z < 0.1$ showing merging features.

   When defining our sample, we excluded AGN hosts with spiral arms when HST images are available.
   We could not do so for AGNs whose images come from the ground-based observation. This can  
 lead to a slight underestimate of the AGN merging fraction because AGN hosts with spiral arms (i.e.,
 a significant disk component) tend to have small $M_{\rm BH}$ values and therefore 
 a smaller merging fraction than those with large $M_{\rm BH}$ values.

  Although we find that 43.6\% of luminous AGNs in our sample show merging features, 
  our simulation of AGNs using the control sample strongly suggests that 
 this is more likely a lower limit of the merging fraction. 
  The true merging fraction of the luminous AGNs can be estimated by using the simulation 
 result. We have shown earlier that the merging fraction is reduced by a factor of 1.7--7.6 at  $z=0.15$ to $z=0.2$ with
 respect to the merging fraction at $z=0.041$. If we make this correction to the observed merging fraction of AGNs, then
 the merging fraction becomes about 75-100\%. 
 Overall, we conclude that the merging is closely related to luminous AGN activity 
 and very much likely to be the triggering mechanism of luminous AGNs.

\section{DISCUSSION\label{discussion}}

  Tidal tails can be more prominent in gas-rich ``wet merging'' \citep{feldmann} 
  and appear from an early stage of galaxy merging \citep{mihos}. 
As the disturbed gaseous and stellar materials
 fall back to the merger product, the tidal tail shrinks and the fall-back material 
can form shell-like or loop-like
 structures \citep{mihos, feldmann}.
 Therefore, we subdivided the merging feature into the tidal-tail type (early stage of merging), 
 the shell type (late stage of merging), and the interaction type 
(very early stage of the merging), 
 as an attempt to see which stage of the merging our AGN hosts are found.
 We find that only one of 17 AGNs with a merging feature shows the shell structure. 
 On the other hand, the majority of AGNs with merging features (15 out of 17) 
 are in the category of the tidal tail type, 
 and the remaining one is the interaction type.
 In comparison, $\sim25\%$ of galaxy mergers in the control sample show 
 the shell structure. 
  At a first glance,
 this result appears to be in perfect agreement with  
 the merger-triggered AGN scenario.
 However, we have shown through our simulation 
 that shell structures are difficult to identify when objects
 are farther away and have a bright central point source. 
  For example,  three of our AGNs are also studied in \cite{bennert}, 
  and they show that two of them 
 (MC 1635+119 and PKS 0736+01) have a shell-type feature in their HST images, 
 but we do not see such a feature in our low-resolution 
 ground-based image, clearly indicating that we are missing the shell-type
 merging features in some cases.
 Therefore,
 we conclude that the lack of tidal-feature-type AGNs in our sample is consistent with a   
 merger-triggered AGN scenario, but a firm conclusion on this issue should be obtained 
 through future studies
 with deeper and higher resolution images of AGNs \citep[e.g.,][]{bennert}.

  We mentioned earlier that we could not identify merging features for some control galaxies
   that have been marked as early types with merging features in \cite{kaviraj}. 
   Consequently, such objects would not be classified to have a merging feature even if they exist in the AGN sample. 
    A correction to this difference in the merging feature classification between classifiers (Kaviraj vs. us) could only more strongly in favor of our result.

  Because some of our AGNs overlap with those studied in previous works, we can make a direct 
 comparison of the classification between ours and the other works. 
 We find that 18 objects in our AGN sample overlap with the AGN hosts studied
 by \cite{Hutchings1992}, \cite{Bahcall}, \cite{mclure}, \cite{dunlop},
 \cite{bennert}, and \cite{letawe}. These studies are based on either HST images
 or ground-based images.  
  We find that three AGNs in the overlap sample are classified to have merging features in
 our study (PKS 1020-103, PKS 2355-082, and PG 1302-102), 
 but they are mentioned only as objects with companions in the other studies,
 demonstrating that our deep images newly uncovered faint merging features in some AGNs.
  On the other hand, we miss two AGNs with a shell-type merging feature as mentioned 
 above, showing that our images are not sensitive to picking up smooth, symmetric, 
 low surface features with detailed structures such as shells.
  One object, PG 0052+251, was classified as Sb in \cite{Bahcall}, but we 
 classify it as a tidal-tail type. The spiral structure is mentioned to be 
 terminated at a companion galaxy in \cite{dunlop}, so the
 spiral structure may be related to a merging activity. The bulge+disk decomposition
 of the host surface-brightness profile 
 indicates that $B/T=0.67$ (K15), suggesting that the host galaxy 
 is an early-type galaxy, lending a farther support that the spiral structure is
 related to the past merging activity. 
  Our classification agrees with that of previous works for  
 the remaining 12 AGNs in the overlap sample.

 Finally, we comment on the contradictory observational results regarding the AGN triggering mechanism.
 Our results that merging features are much more frequently found in luminous AGNs than in quiescent
 galaxies, and that the AGN merging fraction could be as high as 75\%--100\%, point 
 toward a strong connection between merging and luminous AGN activity.  
 Our AGN sample spans
 a range of $10^{44}ergs^{-1}$   $<$ $L_{\mathrm{bol}}$ $<$ $10^{45.4}ergs^{-1}$
 with a median bolometric luminosity of $L_{\mathrm{bol}} \sim 10^{44.52}ergs^{-1}$, 
 and the AGN samples of studies 
 that have not found merging evidence in AGNs extend to a much fainter luminosity range ($L_{\mathrm{bol}} \sim
 10^{42}ergs^{-1}$).  Therefore, the difference in the explored luminosity range could be 
 the main source of the discrepancy, as suggested in previous works such as \cite{Treister}, 
 \cite{hopkins2009}, and \cite{Draper} that major merging is only
 dominant in luminous AGNs. 
 This leads us to a question whether fainter AGNs are really triggered by an internal mechanism.
 After all, SMBHs are predominantly found in bulges of spiral galaxies that are the main hosts of faint
 AGNs.  However, bulges of late-type galaxies can also be produced through merging \citep{Aguerri, Kannappan, Eliche-Moral, hopkins2010, Scannapieco, Oser}.
 Then, if the production of 
 the bulges is closely connected to AGN activity, much like in the case of major 
 merging-triggered AGNs, merging features from merging events (possibly minor merging) could be detectable in less luminous AGNs too.
 The tidal tails from faint satellite galaxies are going to be much less 
prominent than the tidal tails associated with major mergers in luminous
 AGNs, simply if the brightness of tidal tails scales with host galaxy  
 luminosity.  Thus, images that go deeper than current survey limits may reveal 
 traces of the past minor merging activities in less luminous AGNs. 

 For example, studies based on HST survey images have shown little 
 evidence for merging activities in less luminous AGNs. Such imaging data, although very deep in detecting
 point sources, suffer in depth when the surface-brightness limit is considered. This is largely due to 
 the surface brightness dimming effect that scales as $\sim (1+z)^{4}$. 
  In the case of the CANDELS survey (Grogin et al. 2011; Koekemoer et al. 2011) and the images that were  
 used by Schawinski et al. (2012) to draw a conclusion that only 4\% of AGNs at $z \sim 2$ went through
 major merging, the SB limit is only 25.4 AB mag arcsec$^{-2}$ at F775W (similar to $R$ band) when 
 the SB dimming correction is made. 
  Future studies with deeper imaging data will certainly tell us whether 
 the AGN triggering mechanism is closely related to the formation mechanism of 
 bulges. If the classical bulges are formed through merging but the 
 pseudobulges are formed through an internal mechanism (e.g., 
 Kormendy \& Ho 2013), such a difference will be reflected in the existence 
 or nonexistence of merging features around luminous
 AGNs and in the photometric properties of bulges.

\section{SUMMARY}

 To investigate whether galaxy merging can trigger AGN activity or not, 
 we examined deep images (surface-brightness limit of $\mu$ of $\sim$ 27 mag arcsec$^{-2}$) 
 of 39 luminous AGNs at $M_R < -22.6$, $z < 0.3$,  and $7.3 < \mathrm{log} ({M_{\rm BH}/M_\odot}) < 9.0$. We
 find that 17 of 39 AGNs ($\sim 43.6\%$) show evidence of galaxy mergers such as tidal tail, shell, and 
 interaction through careful visual inspection.
 Compared to the control sample of early-type galaxies taken from the SDSS Stripe 82 data at 
 a similar surface-brightness limit, the fraction of AGNs showing the merging feature is about 2.6
 times higher. This difference becomes more significant 
 when we simulate AGNs using the SDSS Stripe 82  early-type galaxy images because our AGNs are located farther away
 than the SDSS Stripe 82 galaxies (surface brightness dimming), and also bright nuclear components of AGNs can make it more difficult to detect
 merging features. Overall, we find that the merging fraction of the simulated AGNs is 
 only 5--15\%, much
 smaller than our result by factors of 4--8 depending on $M_{\rm BH}$ values. 
 Our result strongly suggests that merging plays an important role in triggering AGN activity of luminous AGNs. Future studies with deep imaging observation of fainter AGNs are needed to 
 understand how AGN activities are triggered in fainter AGNs. 

\acknowledgements
This work was supported by the Creative Initiative program, No. 2008-0060544, of the National Research Foundation of Korea (NRFK) funded by the Korean government (MSIP). We thank our CEOU/SNU colleagues for useful discussion and taking some of the data used in this work, and the staffs of the McDonald Observatory, the Maidanak Observatory, and the Las Campanas Observatory for their assistance during the observations. We appreciate useful comments from an anonymous referee. This paper includes the data taken at the McDonald Observatory of The University of Texas at Austin, the Maidanak Observatory, and the Las Campanas Observatory of the Carnegie Institution for Science. MK acknowledges the support from a KASI-Carnegie fellowship. LCH acknowledges support from the Kavli Foundation, Peking University, the Chinese Academy of Sciences, and the Carnegie Institution for Science.

\begin{deluxetable}{lccccccccccccclc}
\tabletypesize{\tiny}
\rotate
\tablecaption{AGN Sample\label{tbl-1}}
\tablewidth{0pt}
\tablehead{
\colhead{Name} & \colhead{$z$} & \colhead{$m_V$} & \colhead{$M_R$} & \colhead{$M_{\mathrm{BH}}$$^{a}$} &
\colhead{Ref.}  & \colhead{$M_{\mathrm{Nuc}}$$^{b}$} &
\colhead{$M_{\mathrm{Bul}}$$^{c}$} & \colhead{$M_{\mathrm{Host}}$$^{d}$} &
\colhead{$\frac{L_{\mathrm{\rm bol}}}{L_{\mathrm{\rm Edd}}}$} & \colhead{Filter} & \colhead{Exp.}& \colhead{$\mu$}& 
\colhead{Seeing}& \colhead{Tel.}& \colhead{Merging}\\
\colhead{} & \colhead{} & \colhead{(mag)} & \colhead{(mag)} & \colhead{($M_\odot$)} &
\colhead{}  & \colhead{(mag)} &
\colhead{(mag)} & \colhead{(mag)} &
\colhead{(erg s$^{-1}$)} & \colhead{} & \colhead{(s)}& \colhead{(mag/${\arcsec}^2$)}& 
\colhead{(arcsec)}& \colhead{}& \colhead{Type}\\
\colhead{(1)} & \colhead{(2)} & \colhead{(3)} & \colhead{(4)} & \colhead{(5)} &
\colhead{(6)} & \colhead{(7)} & \colhead{(8)} &
\colhead{(9)} & \colhead{(10)} &
\colhead{(11)} & \colhead{(12)} & \colhead{(13)}& \colhead{(14)}& 
\colhead{(15)}& \colhead{(16)}
}
\startdata
  3C 206        &    0.2  &    15.76 &    -24.67 &     8.6  &     S(Kim) &      -23.94 &     -22.01 &     -22.01 &     0.204   &     $R$     &     3600  &     26.61 &     0.9   &     D   &     N   \\
  CTS J17.17    &    0.11 &    15.9  &    -23.27 &     8.3  &     -      &       -21.57 &     -22.07 &     -23.27 &     -     &     $i$     &     920   &     26.00 &     0.93  &     C     &     T     \\
  FAIRALL 9     &    0.05 &    13.83 &    -22.86 &     8.2  &     R(Kim) &      -18.97 &     -21.27 &     -21.66 &     0.013  &     $R$     &     1400  &     26.83 &     1.47  &     D   &     T     \\
  HB 890316-346 &    0.27 &    15.1  &    -25.97 &     8.9  &     S(Kim) &    -24.5  &     -22.91 &     -23.21 &     0.043  &     $R$     &     1080  &     25.09 &     0.96  &     D   &     T     \\
  HE 0306-3301  &    0.25 &    15.8  &    -25.27 &     7.9  &     S(Kim) &      -24.63 &     -21.08 &     -22.57 &     0.573  &     $R$     &     2640  &     27.47 &     1.13  &     D   &     N     \\
  HE 0354-5500  &    0.27 &    15.7  &    -25.57 &     7.9  &     S(Kim) &      -24.49 &     -23.31 &     -23.51 &     0.516  &     $R$     &     4800  &     26.10 &     1.12  &     D   &     N     \\
  HE 1434-1600  &    0.14 &    15.5  &    -23.87 &     8.6  &     S(Kim) &     -23.39 &     -22.45 &     -22.45 &     0.040  &     $R$     &     3600  &     27.18 &     0.79  &     D   &     T     \\
  MC 1635+119   &    0.15 &    16.5  &    -22.77 &     7.9  &     S(Kim) &      -21.27 &     -22.49 &     -22.49 &     0.05  &     $R$     &     3600  &     27.57 &     0.92  &     D   &     N     \\
  OX 169        &    0.21 &    15.73 &    -24.67 &     8.5  &     S(Kim) &      -24.5  &     -22.95 &     -23.12  &     0.19  &     $R$     &     3080  &     26.92 &     1.25  &     D   &     T     \\
  PG 0026+12    &    0.15 &    15.41 &    -23.97 &     8.3  &     R(Kim) &      -24.12 &     -22.37 &     -22.37 &     0.261  &     $V$     &     9180  &     26.96 &     1.55  &     Ma &     N     \\
  PG 0052+251   &    0.16 &    15.43 &    -24.47 &     8.3  &     R(Kim) &      -24.12 &     -22.62 &     -23.02  &     0.179  &     $V$     &     9360  &     26.89 &     1.21  &     Ma &     T   \\
  PG 0157+001   &    0.16 &    15.87 &    -23.87 &     7.8  &     S(Kim) &      -22.64 &     -23.65 &     -23.94 &     0.529  &     $V$     &     5400  &     26.11 &     1.15  &     Ma &     T     \\
  PG 0844+349   &    0.06 &    14.5  &    -23.06 &     7.7  &     R(Kim) &      -23.46 &     -21.79 &     -22.11 &     0.412  &     $i$     &     14400 &     27.40 &     2.08  &     Mc&     T     \\
  PG 1004+130   &    0.24 &    15.68 &    -25.07 &     8.9  &     S(Kim) &      -25.59 &     -23.94 &     -23.95 &     0.107  &     $R$     &     3200  &     27.22 &     0.74  &     D   &     N     \\
  PG 1114+445   &    0.14 &    16.12 &    -23.37 &     8.6  &     S(V\&P) &      -22.44 &     -22.68 &     -22.68 &    -     &     $i$     &     10800 &     27.52 &     2.08  &     Mc &     N     \\
  PG 1116+215   &    0.18 &    14.72 &    -25.37 &     8.2  &     S(Kim) &     -24.77 &     -23.29 &     -23.29 &     0.482  &     $R$     &     3600  &     26.65 &     0.95  &     D   &     N     \\
  PG 1211+143   &    0.08 &    14.19 &    -23.97 &     7.9   &     R(Kim) &      -23.84 &     -21.39 &     -21.67 &     0.286  &     $R$     &     3600  &     27.74 &     1.11  &     D   &     N     \\
  PG 1302-102   &    0.28 &    14.92 &    -26.08 &     8.6  &     S(Kim) &      -26.09 &     -23.98 &     -23.99 &     0.403   &     $R$     &     3600  &     26.86 &     0.84  &     D   &     T    \\
  PG 1307+085   &    0.16 &    15.89 &    -23.87 &     8.4  &     R(Kim) &       -24.05 &     -22.25 &     -22.25 &     0.106  &     $R$     &     3600  &     26.90 &     0.93  &     D   &     N    \\
  PG 1309+355   &    0.18 &    15.64 &    -24.67 &     8.4  &     S(Kim) &       -24.63 &     -23.63 &     -23.63 &     0.176  &     $V$     &     10980 &     27.29 &     1.63  &     Ma &     N    \\
  PG 1322+659   &    0.17 &    15.84 &    -24.07 &     8.3  &     S(V\&P) &     -23.83 &     -22.06 &     -22.06 &     -    &     $V$     &     11100 &     27.75 &     1.83  &     Ma &     N     \\
  PG 1351+64    &    0.09 &    14.28 &    -23.97 &     8.5  &     S(Kim) &      -23.44 &     -22.18 &     -22.27 &     0.075  &     $V$     &     9720  &     27.41 &     1.91  &     Ma &     N     \\
  PG 1352+183   &    0.15 &    16.68 &    -22.97 &     8.4  &     S(V\&P) &      -21.88 &     -22.34 &     -22.34 &     -    &     $V$     &     10800 &     27.15 &     1.86  &     Ma &     N     \\
  PG 1402+261   &    0.17 &    15.34 &    -24.57 &     7.5  &     S(Kim) &      -23.23 &     -21.17 &     -22.51 &     0.486  &     $V$     &     12960 &     27.06 &     1.47  &     Ma &     N     \\
  PG 1411+442   &    0.09 &    14.01 &    -24.57 &     8.4  &     R(Kim) &      -23.6  &     -21.54 &     -22.3 &     0.076  &     $i$     &     12600 &     27.40 &     2.54  &     Mc &     T     \\
  PG 1416-129   &    0.13 &    16.1  &    -22.97 &     8.5  &     S(Kim) &       -22.82 &     -21.59 &     -21.59 &     0.044  &     $R$     &     2400  &     27.37 &     0.92  &     D   &     N    \\
  PG 1440+356   &    0.08 &    14.58 &    -23.37 &     7.4  &     S(Kim) &        -23.19 &     -20.91 &     -22.17 &     0.608  &     $V$     &     7550  &     27.52 &     1.17  &     Ma &     T   \\
  PG 1519+226   &    0.14 &    16.5  &    -22.97 &     7.9  &     S(V\&P) &      -22.66 &     -21.38 &     -21.38 &     -    &     $i$     &     9000  &     27.57 &     2.08  &     Mc &     N     \\
  PG 1613+658   &    0.13 &    15.49 &    -23.37 &     8.2  &     R(Kim) &      -24.32 &     -23.08 &     -24.08 &     0.297   &     $i$     &     5400  &     26.18  &     2.02  &     Mc &     T    \\
  PG 1617+175   &    0.11 &    15.39 &    -23.37 &     8.7  &     R(Kim) &       -23.22 &     -21.21 &     -21.38 &     0.029  &     $R$     &     3600  &     27.41 &     0.88  &     D   &     N    \\
  PG 1626+554   &    0.13 &    15.68 &    -23.67 &     8.5   &     S(V\&P) &      -23.22 &     -22.5  &     -22.5 &     -    &     $i$     &     6000  &     27.32 &     2.54  &     Mc &     N     \\
  PG 2214+139   &    0.07 &    14.66 &    -22.96 &     8.6  &     S(V\&P) &       -21.56 &     -22.61 &     -22.61 &     -    &     $V$     &     10800 &     27.36 &     1.1   &     Ma &     S     \\
  PKS 0159-062  &    0.19 &    16.6  &    -23.17 &     8.7   &     -      &       -20.94 &     -22.87 &     -22.9 &      -    &     $r$     &     2180  &     27.96 &     0.73  &     C     &     T     \\
  PKS 0736+01   &    0.19 &    16.47 &    -23.47 &     8.1  &     S(Kim) &       -24.12 &     -23.39 &     -23.39 &     0.381  &     $R$     &     3600  &     27.52 &     1.48  &     D   &     N    \\
  PKS 1020-103  &    0.2  &    16.11 &    -24.17 &     8.7  &     S(Kim) &        -23.38 &     -23.02 &     -23.02  &     0.061  &     $R$     &     3600  &     27.20 &     1.19  &     D   &     I  \\
  PKS 1217+02   &    0.24 &    15.97 &    -24.77 &     8.3  &     S(Kim) &      -24.28 &     -23.4  &     -23.4 &     0.278  &     $R$     &     3600  &     27.02 &     0.74  &     D   &     N     \\
  PKS 2135-14   &    0.2  &    15.53 &    -24.87 &     8.9  &     S(Kim) &        -23.98 &     -23.12 &     -23.12  &     0.043  &     $R$     &     4440  &     26.48 &     1.57  &     D   &     N  \\
  PKS 2349-01   &    0.17 &    16.59 &    -23.27 &     8.6  &     S(Kim) &       -23.81 &     -23.34 &     -23.75 &     0.12  &     $R$     &     3080  &     27.25 &     0.84  &     D   &     T     \\
  PKS 2355-082  &    0.21 &    17.5  &    -23.07 &     8.6  &     S(Kim) &      -23.02 &     -23.32 &     -23.5  &     0.034  &     $R$     &     4680  &     26.85 &     0.94  &     D   &     T     \\
\enddata

\tablecomments{Column 1: object name;
 column 2: redshift;
 column 3: apparent magnitude in the $V$ band from \cite{veron}; 
 column 4; absolute $R$-band magnitude, converted from absolute $B$-band magnitude;
 column 5: the black hole mass;
 column 6: references for M$_{\rm BH}$ and method for estimating M$_{\rm BH}$: R=reverberation mapping; S=single-epoch method; Kim=\cite{kim2008a}; K15; V\&P=\cite{vp}; 
 column 7: absolute $R$-band magnitude of the AGN nuclear component;
 column 8: absolute $R$-band magnitude of the bulge of the AGN;
 column 9: absolute $R$-band magnitude of the AGN host galaxy (see section \ref{mag_agn} for how we derived quantities in columns 5 and 7--9). Bulge magnitudes come from an SB fitting of the images from D or C in the 
 column 15 and from the black hole mass for Ma or Mc in the column 15;
 column 10: Eddington ratio from \cite{kim2008a};  
 column 11: filter; 
 column 12: integrated exposure time;
 column 13: surface-brightness limit, averaged over a 1$\arcsec$ $\times$ 1 $\arcsec$ area 
 at 1$\sigma$ per pixel in the observed band;  
 column 14: seeing FWHM in arcseconds;
 column 15: telescope: D=2.5m DuPont telescope in Las Campanas, C=CFHT, Ma=1.5m telescope in Maidanak observatory, Mc=2.1m telescope in McDonald observatory;
 column 16: merging type T=tidal tail; S=shell structures; I=interaction; N=no merging feature.\\
 $^{a}$ Typical formal error is $<0.1$ dex, but the error arising from the intrinsic scatter of the
 $M_{\rm BH}$ estimators could be as much as 0.4 dex. \\
 $^{b}$Typical error is $\sim0.1$ mag. \\
 $^{c}$Typical error is $\pm0.3$ -- $\pm0.4$ mag. \\
 $^{d}$Typical error is $\pm0.2$. 
   }
\end{deluxetable}
\clearpage

\begin{deluxetable}{lccccccccc}
\tabletypesize{\scriptsize}
\tablecaption{Fitted Parameters\label{tbl-2}}
\tablewidth{0pt}
\tablehead{
\colhead{Name} & \colhead{$m_{\mathrm{Nuc}}$} & \colhead{$m_{\mathrm{Bul}}$} & \colhead{$R_{\mathrm{e}}$} & \colhead{$n$} &\colhead{$m_{\mathrm{disk}}$}& \colhead{$R_{\mathrm{d}}$} & \colhead{$n_{\mathrm{d}}$}
&\colhead{${\chi_{\upsilon}}^2$} & \colhead{Filter}\\
\colhead{} & \colhead{(mag)}& \colhead{(mag)}& \colhead{($\arcsec$)}& \colhead{} & \colhead{(mag)}& \colhead{[$\arcsec$]} &\colhead{}&\colhead{}&\colhead{}\\
\colhead{(1)}& \colhead{(2)}& \colhead{(3)}& \colhead{(4)}& \colhead{(5)}& \colhead{(6)}& \colhead{(7)}& \colhead{(8)}& \colhead{(9)}& \colhead{(10)}
}
\startdata
CTS J17.17 & 16.74 $\pm$ 0.00 & 16.24 $\pm$ 0.01 & 2.2 $\pm$ 0.01 &  1.27 $\pm$ 0.02 & 16.4 $\pm$ 0.01 & 6.23 $\pm$ 0.16 & 1 & 1.18 &  i\\
PKS 0159-062 & 19.11 $\pm$ 0.01 & 17.37 $\pm$ 0.00 & 4.36 $\pm$ 0.03 & [4] & - & - & - & 1.11 & r\\			
\enddata
\tablecomments{ column 1: object name of CFHT data fitted through GALFIT; 
 column 2: apparent nuclear magnitude in the observed filter; 
 column 3: apparent bulge magnitude in the observed filter;
 column 4: effective radius of the bulge;
 column 5: Sersic index of the bulge;  
 column 6: apparent disk magnitude in the observed filter;
 column 7: disk scale length;
 column 8: sersic index of the disk;
 column 9: reduced $\chi^{2}$;
 column 10: filter.}
\end{deluxetable}

\clearpage

\begin{deluxetable}{rccccccccccccc}
\tabletypesize{\tiny}
\rotate
\tablecaption{Control Sample of Early-type Galaxies\label{tbl-control}}
\tablewidth{0pt}
\tablehead{
\colhead{Sequence} & \colhead{SDSS ID} & \colhead{R.A.} & \colhead{Decl.} & \colhead{$z$} &
\colhead{$R$}  & \colhead{Error} &
\colhead{$B/T$} & \colhead{Error} &
\colhead{$M_{R}$} & \colhead{Error} & \colhead{$M_{\rm BH}$}& \colhead{Type}& 
\colhead{Type} \\
\colhead{(Number)} & \colhead{} & \colhead{(deg)} & \colhead{(deg)} & \colhead{} &
\colhead{(mag)}  & \colhead{(mag)} &
\colhead{} & \colhead{} &
\colhead{(mag)} & \colhead{(mag)} & \colhead{log$(M/M_{\odot})$} & \colhead{Kaviraj (2010)}& 
\colhead{This work} \\
\colhead{(1)} & \colhead{(2)} & \colhead{(3)} & \colhead{(4)} & \colhead{(5)} &
\colhead{(6)} & \colhead{(7)} & \colhead{(8)} &
\colhead{(9)} & \colhead{(10)} &
\colhead{(11)} & \colhead{(12)} & \colhead{(13)}& \colhead{(14)}
}
\startdata
    2 & 587730845817504263 & 321.915192 & -1.102175 &    0.030 &   14.739 &    0.003 &    0.150 &    0.000 &  -18.800 &    0.020 &    6.875 & 1 & N \\
   10 & 587730846891507736 & 322.370728 & -0.219635 &    0.050 &   14.415 &    0.002 &    0.410 &    0.010 &  -21.320 &    0.020 &    8.135 & 4 & T \\
   13 & 587730846891507870 & 322.451935 & -0.313562 &    0.030 &   14.518 &    0.002 &    0.550 &    0.000 &  -20.490 &    0.010 &    7.720 & 3 & N \\
   24 & 587730847963939362 & 319.503601 & 0.570868 &    0.035 &   15.062 &    0.003 &    0.940 &    0.020 &  -20.980 &    0.020 &    7.965 & 1 & N \\
   49 & 587730848501596412 & 321.287964 & 0.896778 &    0.049 &   15.267 &    0.003 &    0.490 &    0.010 &  -20.660 &    0.010 &    7.805 & 1 & N \\
   50 & 587730848501399575 & 320.744537 & 1.026719 &    0.031 &   14.173 &    0.002 &    0.500 &    0.010 &  -20.630 &    0.010 &    7.790 & 1 & N \\
   76 & 587731174382567703 & 320.190460 & 0.790999 &    0.043 &   15.405 &    0.003 &    0.990 &    0.010 &  -21.080 &    0.010 &    8.015 & 1 & N \\
   89 & 587731185116053719 & 327.825348 & -0.957031 &    0.027 &   13.665 &    0.002 &    1.000 &  -99.990 &  -21.584 &    0.002 &    8.267 & 2 & T \\
  112 & 587731185125818492 & 350.121124 & -1.002414 &    0.031 &   13.519 &    0.002 &    1.000 &  -99.990 &  -22.015 &    0.002 &    8.483 & 1 & N \\
  128 & 587731185650565143 & 322.302155 & -0.621743 &    0.030 &   14.433 &    0.002 &    0.520 &    0.000 &  -20.360 &    0.010 &    7.655 & 2 & T \\
  129 & 587731185650565167 & 322.305756 & -0.523884 &    0.030 &   14.723 &    0.003 &    0.530 &    0.000 &  -20.060 &    0.010 &    7.505 & 1 & N \\
  135 & 587731185651155402 & 323.751648 & -0.511436 &    0.030 &   14.439 &    0.002 &    0.500 &    0.000 &  -20.270 &    0.010 &    7.610 & 2 & T \\
  146 & 587731185654366563 & 331.085999 & -0.502819 &    0.046 &   15.855 &    0.004 &    0.340 &    0.010 &  -19.310 &    0.040 &    7.130 & 1 & N \\
  192 & 587730846893080588 & 325.963776 & -0.321398 &    0.027 &   14.169 &    0.002 &    0.570 &    0.010 &  -20.400 &    0.010 &    7.675 & 1 & N \\
  221 & 587731186724110554 & 321.939697 & 0.326403 &    0.031 &   14.692 &    0.003 &    0.750 &    0.010 &  -20.540 &    0.020 &    7.745 & 1 & N \\
  237 & 587731186735186207 & 347.261871 & 0.266889 &    0.033 &   15.056 &    0.003 &    0.560 &    0.010 &  -19.680 &    0.020 &    7.315 & 2 & T \\
  261 & 587731187271663782 & 346.281830 & 0.826253 &    0.042 &   15.627 &    0.003 &    0.960 &    0.060 &  -20.980 &    0.070 &    7.965 & 2 & T \\
  266 & 587731187279069315 & 3.300832 & 0.742940 &    0.039 &   14.342 &    0.002 &    0.880 &    0.010 &  -21.640 &    0.010 &    8.295 & 1 & N \\
  269 & 587731187279593490 & 4.393614 & 0.743043 &    0.044 &   14.596 &    0.002 &    0.880 &    0.010 &  -21.790 &    0.010 &    8.370 & 2 & T \\
  272 & 587731187279462423 & 4.088767 & 0.788624 &    0.044 &   15.253 &    0.003 &    0.700 &    0.010 &  -20.790 &    0.010 &    7.870 & 1 & N 
\enddata
\tablecomments{column 1: sequence number of Kaviraj (2010);   
column 2: SDSS object ID;  
column 3: R.A. (J2000); 
column 4: Decl. (J2000); 
column 5: redshift; 
column 6-7: apparent $R$-band magnitude and its error; 
column 8-9: $B/T$ and its error. The error of -99.990 indicates that the SB model does not have a disk component; 
column 10-11: absolute $R$-band magnitude of the bulge component and its error; 
column 12: black hole mass; 
column 13: morphological classification from Kaviraj (2010): 1:relaxed, 2:tidal feature, 3:dust feature, 4:tidal and dust feature, 5:interacting, 7:spheroids with faint disk; 
column 14: morphological classification from this work: N:relaxed, T:tidal feature, S:shell-type.     
(This table is available in its entirety in a machine-readable form in the online journal. A portion is shown here 
for guidance regarding its form and content.)  
  }
\end{deluxetable}
\clearpage

\begin{deluxetable}{lcccccccccccc}
\tabletypesize{\scriptsize}
\tablecaption{Merging Fraction of Simulated AGNs I\label{tbl-3}}
\tablewidth{0pt} 
\tablehead{
\colhead{} &  \multicolumn{5}{c}{log$(M_{\rm BH}/M_{\odot}) < 8.39$} &\colhead{} & 
\multicolumn{5}{c}{log$(M_{\rm BH}/M_{\odot}) \geq 8.39$}\\
\cline{2-6}  \cline{8-12}\\
\colhead{} & \multicolumn{11}{c}{Redshift}\\
\colhead{$m_{\mathrm{Nuc}}$} & \colhead{0.041} & \colhead{0.086} & \colhead{0.138} & \colhead{0.2} &\colhead{0.275}
&\colhead{} & \colhead{0.041} & \colhead{0.086} & \colhead{0.138} & \colhead{0.2} &\colhead{0.275}\\
}
\startdata
 None &  0.141 & 0.116 & 0.074 & 0.0141 & 0.011 && 0.455 & 0.394 & 0.303 & 0.091 & 0.061\\
 $m_{\mathrm{gal}}$ & 0.137 & 0.106 & 0.067 & 0.014 & 0.007 && 0.424 & 0.364 & 0.242 & 0.091 & 0.061 \\		
 $m_{\mathrm{gal}}$ -1.0  & 0.127 & 0.095 & 0.042 & 0.004 & 0.004 && 0.394 & 0.303 & 0.212 & 0.061 & 0.030\\
 $m_{\mathrm{gal}}$ -1.5 & 0.123 & 0.085 & 0.042 & 0.004 & 0.004 && 0.364 & 0.242 & 0.212 & 0.061 & 0.030\\
\enddata

\tablecomments{
The simulated AGNs are divided into two $M_{\rm BH}$ bins, and  
the results are for the seeing of 1$\arcsec$. They are simulated to
be at several different redshifts and to have various nuclear magnitudes.
  }
\end{deluxetable}

\begin{deluxetable}{lcccccccccccc}
\tabletypesize{\scriptsize}
\tablecaption{Merging Fraction of Simulated AGNs II\label{tbl-4}}
\tablewidth{0pt} 
\tablehead{
\colhead{} &  \multicolumn{5}{c}{1 $ \arcsec$} &\colhead{} & \multicolumn{5}{c}{2 $\arcsec$}\\
\cline{2-6}  \cline{8-12}\\
\colhead{} & \multicolumn{11}{c}{Redshift}\\
\colhead{$m_{\mathrm{Nuc}}$} & \colhead{0.041} & \colhead{0.086} & \colhead{0.138} & \colhead{0.2} &\colhead{0.275}
&\colhead{} & \colhead{0.041} & \colhead{0.086} & \colhead{0.138} & \colhead{0.2} &\colhead{0.275}\\
}
\startdata
None & 0.167  &   0.142  &   0.098   &  0.022  &  0.016  && 0.164  &   0.132  &   0.069   &  0.009  &  0.003\\
$m_{\mathrm{gal}}$ & 0.167   &  0.132  &  0.085  &  0.022  &  0.013  && 0.155   &  0.12  &  0.057  &  0.009  &  0.003\\		
$m_{\mathrm{gal}}$ -1.0  & 0.151  &  0.11  &  0.06  &  0.009  &  0.006 & & 0.145  &  0.085  &  0.041  &  0.003  &  0.00\\
$m_{\mathrm{gal}}$ -1.5 & 0.151  &  0.11  &  0.06  &  0.009 &  0.006  && 0.145  &  0.085  &  0.038  &  0.003 &  0.00	\\
\enddata

\tablecomments{The results for two
seeing conditions (1$\arcsec$ and 2$\arcsec$) are presented for simulated AGNs placed at
several different redshifts and with different nuclear magnitudes.  
  }
\end{deluxetable}

\begin{deluxetable}{lccccc}
\tabletypesize{\scriptsize}
\tablecaption{Merging Fraction of AGNs versus Simulated AGNs\label{table_mf}}
\tablewidth{0pt} 
\tablehead{
\colhead{} &  \multicolumn{2}{c}{AGN} & \colhead{} &\colhead{Simulated AGN} \\
\cline{2-3} \cline{5-5}\\
\colhead{log$(M_{\rm BH}/M_{\odot})$} & \colhead{Base} & \colhead{Best} & \colhead{} & \colhead{$m_{\rm PSF} = m_{gal}-1$}\\
\colhead{} & \colhead{}& \colhead{($M_{\mathrm{Nuc}}(R) < -22.44$)}& \colhead{}& \colhead{1$\arcsec$ seeing}
}
\startdata
 $7.3 <$ log$(M_{\rm BH}/M_{\odot})< 8.39$ & 0.389$^{+0.121}_{-0.098}$   &  0.333 $^{+0.135}_{-0.096}$  &&  0.051 $^{+0.017}_{-0.010}$   \\		
&&&\\		
 $8.39 \leq$ log$(M_{\rm BH}/M_{\odot}) < 9.0$ & 0.476 $^{+0.106}_{-0.102}$   &  0.444 $^{+0.117}_{-0.106}$  &&  0.117 $^{+0.083}_{-0.036}$   \\				
\enddata

\end{deluxetable}

\clearpage

\clearpage

\clearpage

\clearpage

\clearpage

\clearpage

\clearpage

\clearpage

\clearpage

\clearpage

\clearpage

\clearpage

\clearpage

\clearpage

\clearpage

\begin{figure}
\includegraphics[width=16cm]{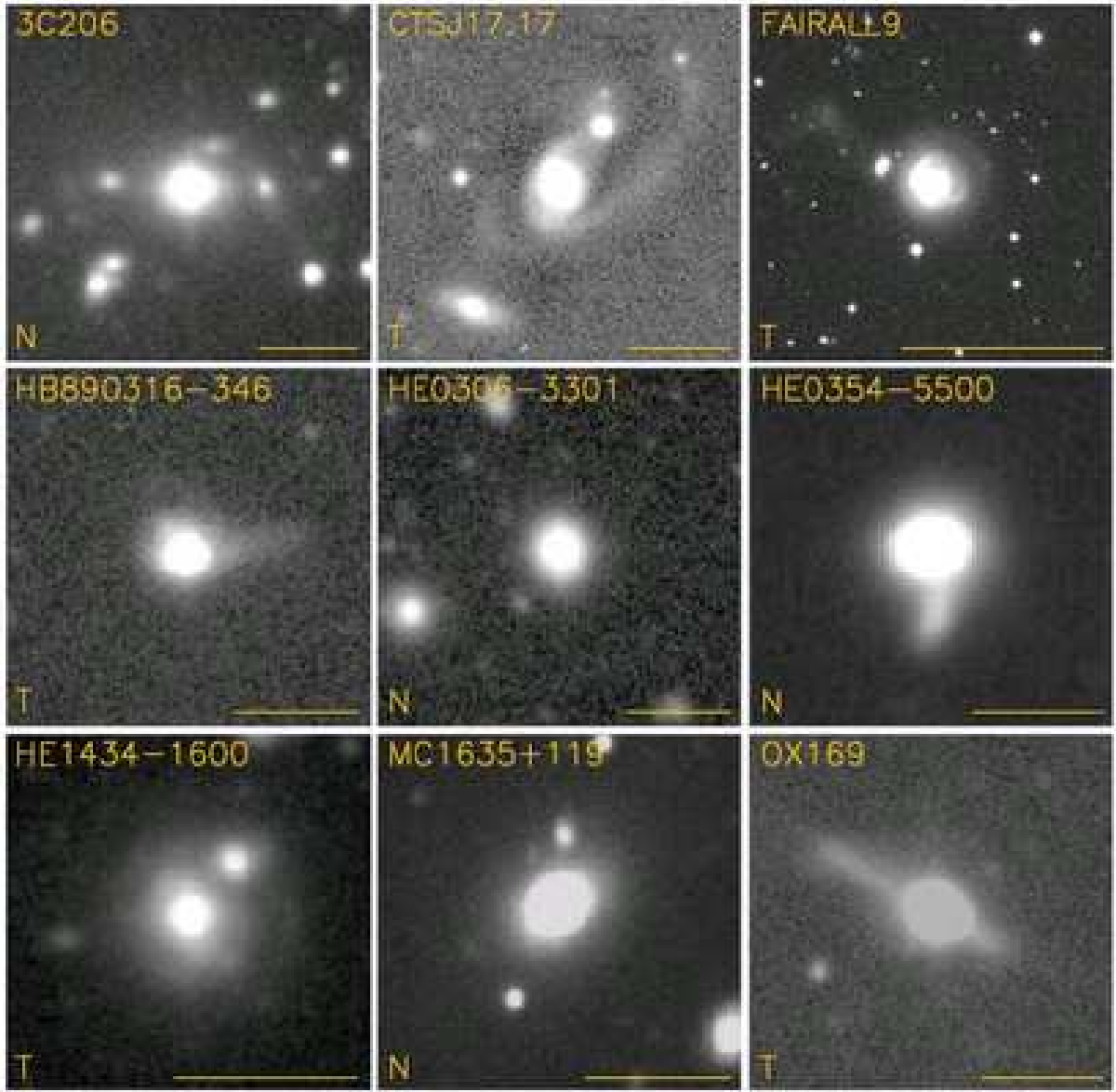}
\caption{Images of 39 AGNs in our base sample. T=tidal-tail type, 
S=shell type, I=Interaction type. The length of horizontal bar in each panel is 10$\arcsec$.}
\label{merging1}
\end{figure}
\clearpage

\begin{figure}
\figurenum{16}
\includegraphics[width=16cm]{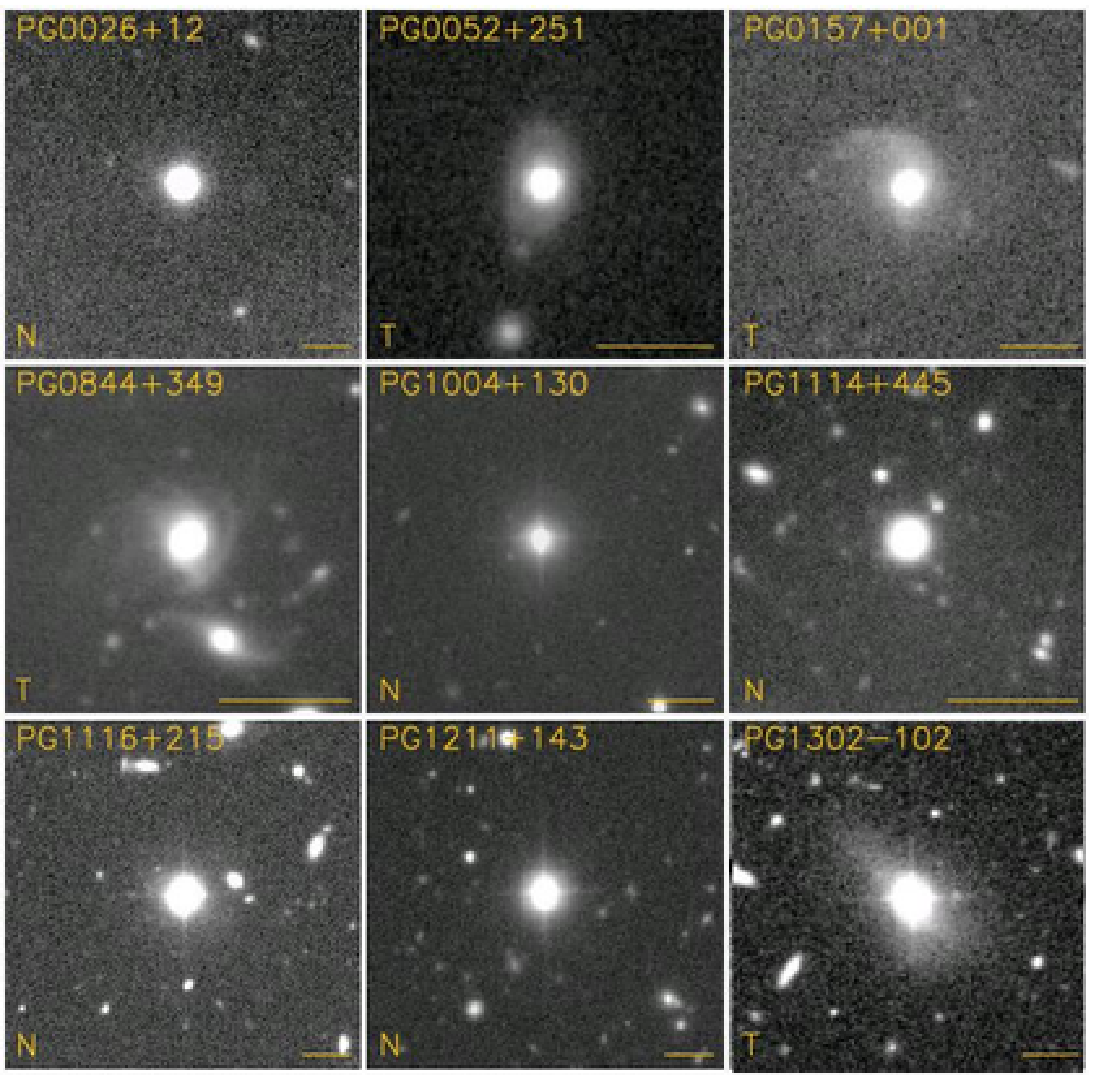}
\caption{Continued}\label{merging1}
\end{figure}
\clearpage

\begin{figure}[p]
\figurenum{16}
\includegraphics[width=16cm]{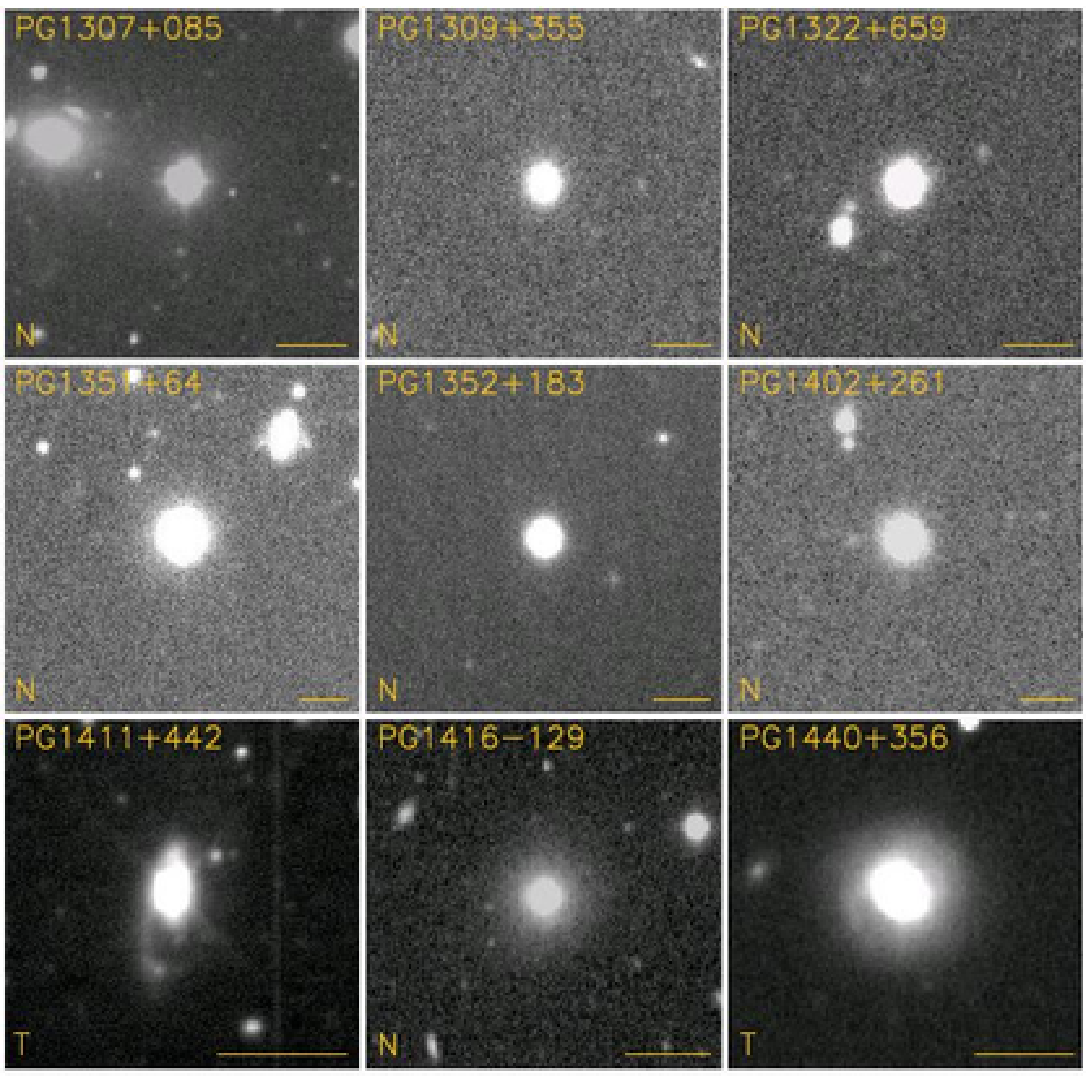}
\caption{Continued}\label{merging1}
\end{figure}
\clearpage

\begin{figure}[p]
\figurenum{16}
\includegraphics[width=16cm]{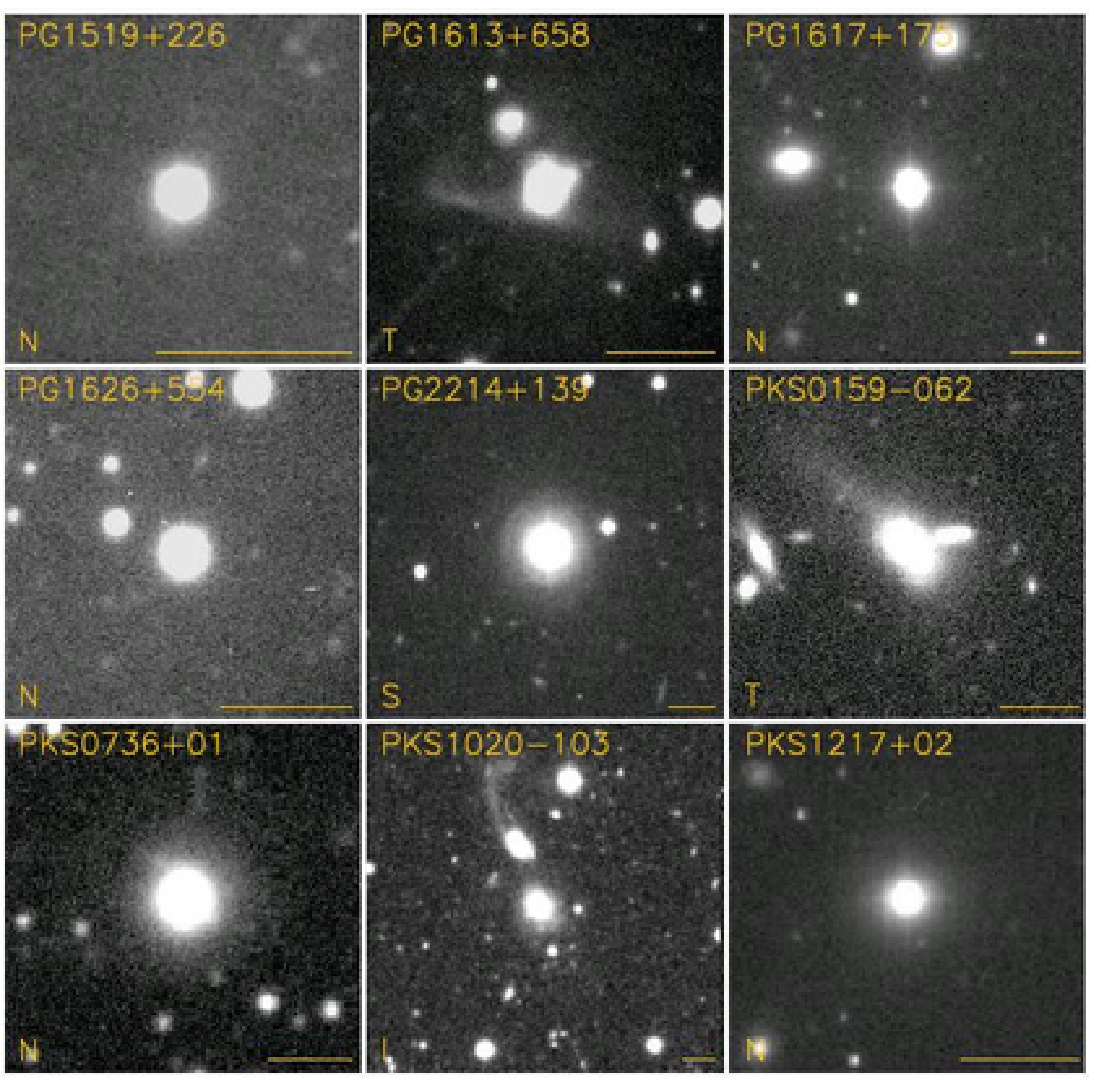}
\caption{Continued}\label{merging1}
\end{figure}
\clearpage

\begin{figure}[p]
\figurenum{16}
\includegraphics[width=16cm]{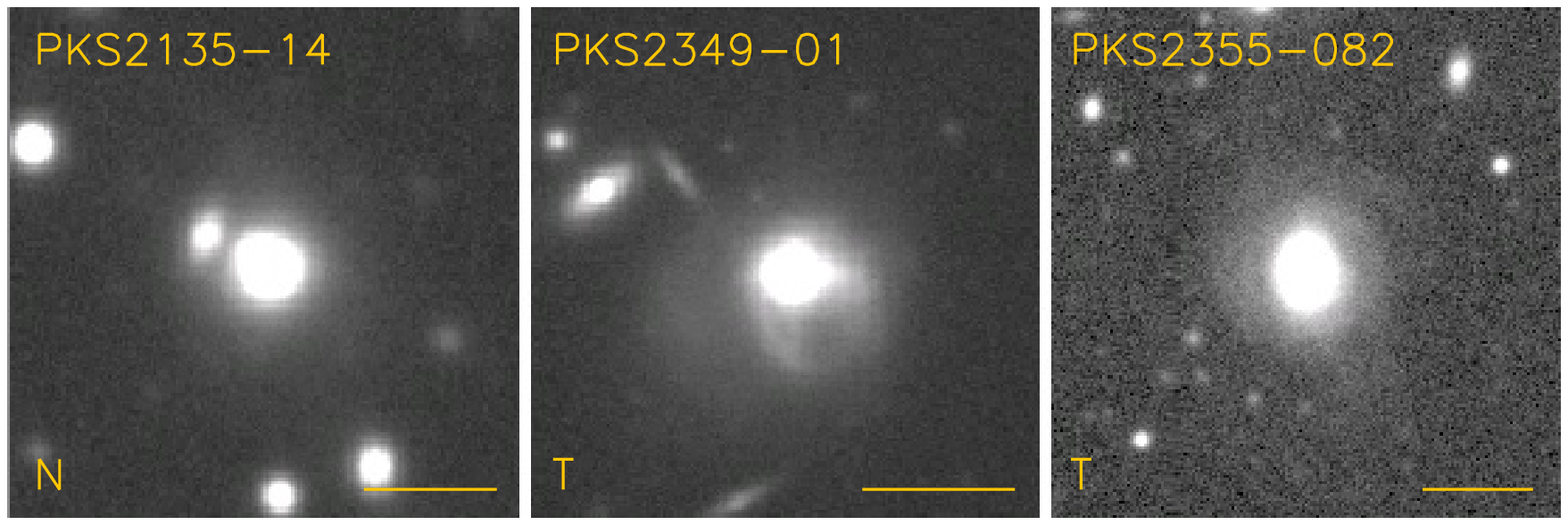}
\caption{Continued}\label{merging1}
\end{figure}
\clearpage

 \begin{figure}[p]
\includegraphics[width=16cm]{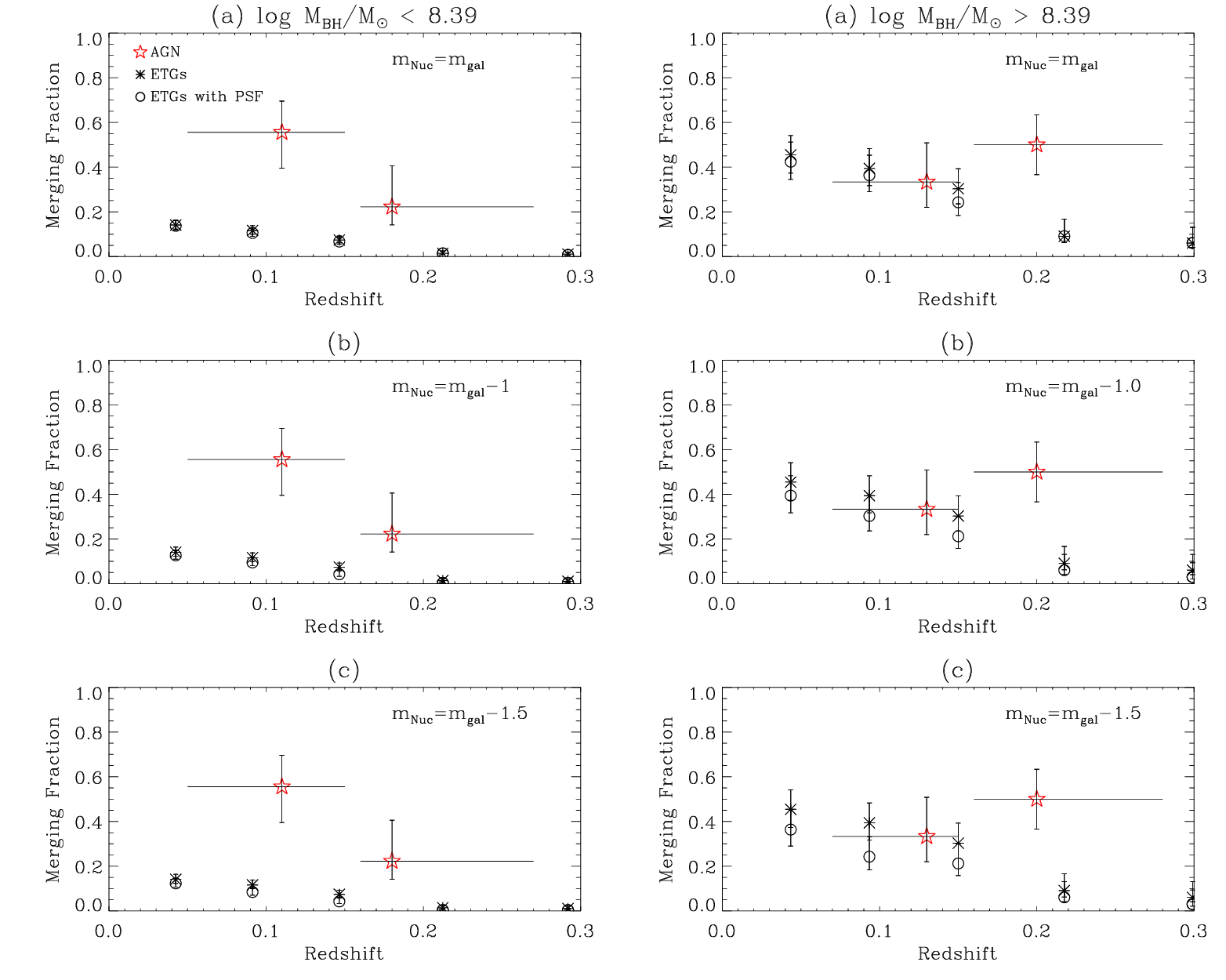}
\caption{Merging fraction of AGNs and the simulated AGNs as a function of redshift. 
 The left panel is the merging fraction of objects with $\mathrm{log (M_{BH}/M_{\odot})} < 8.39$,
 and the right panel is the merging fraction of objects with $\mathrm{log (M_{BH}/M_{\odot})} \geq 8.39$.
  The panels for each $M_{\mathrm{BH}}$ are also divided into three, each
  corresponding to a different nuclear magnitude.
  The magnitudes of the added PSF components for each panel are (a) top: $m_{\mathrm{gal}}$, (b) middle: $m_{\mathrm{gal}}-1.0$, and (c) bottom: $m_{\mathrm{gal}}-1.5$. 
  The black asterisks are the merging fraction of early-type galaxies without any nuclear source (PSF)
 included, and
 the black circles are the numbers for the cases with a nuclear component included. 
 The seeing of simulated images is assumed to be 1$\arcsec$ here.}\label{mf_z}
\end{figure}
\clearpage

\begin{figure}[p]
\includegraphics[width=14cm]{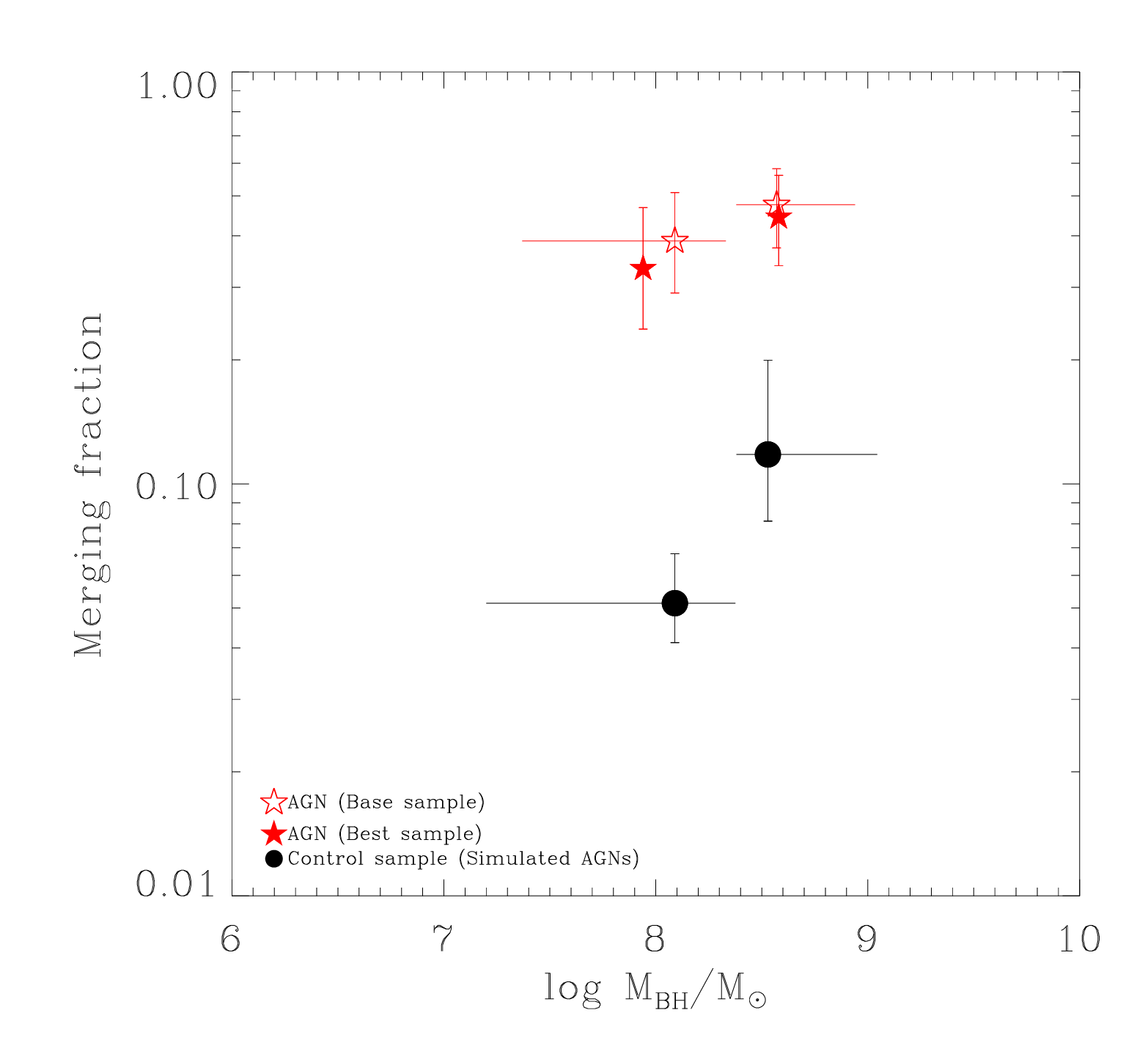}
\caption{Merging fraction of AGNs and simulated AGNs (control sample) 
 versus the black hole mass of the host galaxy. The merging fraction of simulated AGNs has 
 been corrected for the difference in the $M_{\rm BH}$ distribution between the AGN and the control
 samples.
 The red stars are for AGNs in the base sample, and the filled red stars are for 
 AGNs in the best sample ($M_{\mathrm{Nuc}}(R) < -22.44$). The filled black circles are for simulated AGNs.}\label{mf_bh_sum}
\end{figure} 
\clearpage

\begin{figure}[p]
\includegraphics[width=14cm]{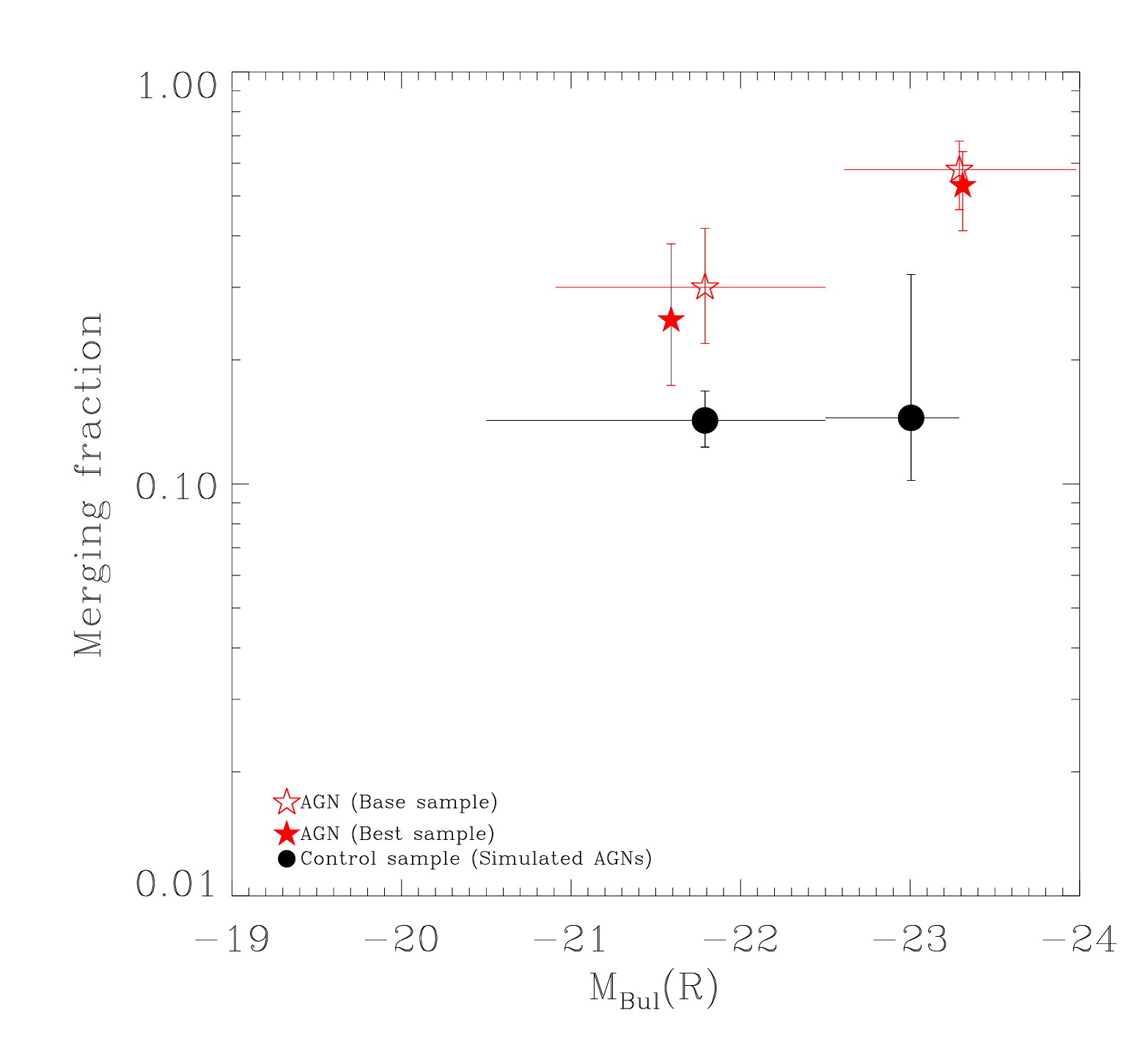}
\caption{Merging fraction of AGNs and simulated AGNs (control sample) versus the bulge magnitude of the host galaxy. The merging fraction of simulated AGNs has 
 been corrected for the difference in the $M_{R}$ distribution between the AGN and the control
 samples.
The red stars are for AGNs in the base sample, the filled red stars are for 
 AGNs in the best sample ($M_{\mathrm{Nuc}}(R) < -22.44$), and the filled black circles are for simulated AGNs.}\label{mf_mbul_sum}
\end{figure}
\clearpage

\end{document}